# An End-User Development approach for Mobile Web Augmentation

Gabriela Bosetti [1], Sergio Firmenich [1,2]
Silvia E. Gordillo [1,3], Gustavo Rossi [1,2] and Marco Winckler [4]

[1] LIFIA. Facultad de Informática. UNLP. La Plata, Argentina
```
{ gabriela.bosetti, sergio.firmenich,
gordillo, gustavo}@lifia.info.unlp.edu.ar
```
[2] Consejo Nacional de Investigaciones Científicas y Técnicas. Argentina
[3] Comisión de Investigaciones Científicas. Argentina
[3] ICS-IRIT, University of Toulouse 3. France
```
winckler@irit.fr
```

**Abstract.** The trend towards mobile devices usage has put more than ever the Web as a ubiquitous platform where users perform all kind of tasks. In some cases, users access the Web with "native" mobile applications developed for well-known sites, such as LinkedIn, Facebook, Twitter, etc. These native applications might offer further (e.g. location-based) functionalities to their users in comparison with their corresponding Web sites, because they were developed with mobile features in mind. However, most Web applications have not this native mobile counterpart and users access them using browsers in the mobile device. Users might eventually want to add mobile features on these Web sites even though those features were not supported originally. In this paper we present a novel approach to allow end users to augment their preferred Web sites with *mobile features.* This end-user approach is supported by a framework for mobile Web augmentation that we describe in the paper. We also present a set of supporting tools and a validation experiment with end users.

## 1. Introduction

The increasing growth of both the Web and mobile technology in today's global scenario has raised new forms of participation of end users. Mobile devices started to be used as personal computers, competing and even replacing computers for daily life tasks and, as far as the resources of these devices increased, so it did –and still does– the number of innovative implemented functionalities they can benefit from. Today, it is usual that end users interact with the same Web applications from both, desktop computers and mobile devices.

Some of the most popular Web applications (like Facebook, Twitter, YouTube and others) provide native mobile applications, and some of them (like Booking) also provide nice location-based features to profit from the mobile nature of user devices. However, the majority of Web sites are still accessed from Web browsers in the device, and they do not provide any mobile features. This phenomenon might be caused either by economic reasons (to build the native application), by the difficulties to modify the Web application to support the mobile features or just for lack of interest. In any case, end users lack the possibility to a better access to the information and services provided by these applications.

This situation is worse for Web applications providing information that is naturally location-based, such as the one presented by museums, city halls or tourist applications. In these cases, accessing parts of such information "in-situ" (e.g. visiting the museum with a smart-phone) could certainly enrich the visit, providing the end user with locative media experiences, as it has been shown in dozens of cases [12; 37]. Some of these institutions provide a "poor man" location-based feature by adding QR-codes to some of the points of interest (artworks, monuments) so users can explore some information with their devices (e.g. using QRPedia[1]). More complex scenarios, like itinerant or temporary exhibitions, might complicate things further. In this paper we present an approach which aims to empower end users to implement mobile Web applications by profiting from information (and services) already existing in the original Web sites, but enriching them with different kinds of mobile functionality. The underlying philosophy of our approach is the one provided by the concept of Web augmentation.

---
[1] QRPedia: http://qrpedia.org/

During the last years, end users started to –unofficially– add new functionalities to Web applications when some of their requirements were not contemplated originally. In this way, users began to participate not only under the role of *consumers* of Web applications, but also learned how to become *producers* of their own solutions. Being a producer is not necessarily synonymous of having –or not– technical knowledge or expertise. Within this category of end users, you can find people with skills for textual or visual programming, as well as those who do not have much technical knowledge but can also build their applications by using simple or assisted tools, as form-based wizards or tools supporting programming by example [29].

A very popular technique for adapting existing third-party Web applications is Web Augmentation [4; 17]. There are different strategies for achieving Web Augmentation (WA); one of them is client-side scripting, which consists in manipulating the applications' user interface (UI) when a particular site is already loaded on the browser. With this technique, style, structure and functionality can be added, modified or removed from a Web page without the need of changing the source code at server-side. Such scripts that perform a specific adaptation, are called *augmenters* [17], and it is usual that their creators have some level of expertise in JavaScript.

The amount of features that can be opportunistically added to an existing application are countless, moreover taking into account the possibility of consuming information from a wide range of sources. For instance, EBay products can be augmented with more information for the user to decide whether to purchase it or not by summarizing opinions on the sellers, checking the price of the product in other sites, etc.

In this paper, we focus on adding different kinds of mobile features to Web applications by using scripting. Applying Web augmentation on mobile devices implies that, besides the common aspects that may be adapted typically (e.g. look and feel, personalization, recommendations, etc), mobile and context-aware features can be contemplated as well. For instance, we could take into account the user's position for augmenting a News portal with geo-positioned and content-related videos and tweets. By taking the current user's position through the Geoposition Web API[2], it is possible to use it for building geo-located queries through the Data API of Youtube[3] and the REST API of Twitter[4]. Then, the retrieved content could be injected into a News portal's Web page by using the Addon-SDK[5] of Firefox. We can listen for device's orientation changes (DeviceOrientationEvent[6]) and augment Google Maps' Web page with a real-world view (camera capture) and an arrow pointing to the target location when the user tilts the device vertically. We can calculate the perceived sound level by using the WebRTC API[7] for automatically adapting the volume of Youtube videos at middle levels, or stopping/resuming the reproduction at high ones.

Although there are already some approaches for augmenting Web applications from mobile Web browsers, such as [7; 23; 40], they are aimed and limited to *producers with* programming skills (from now on, *developers*).

We have developed the Mobile Web Augmentation (MoWA) approach [3; 8], which comprises a software framework and a set of tools for developers. Basically, MoWA provides developers with a framework for creating mobile Web applications based on client-side adaptation including the addition of new (e.g. location-based) contents and functionality (such as context-awareness) directly on the front-end, i.e., the Web browser. In this paper, we change our target audience from developers to a broader one: producers with no programming skills (from now on just *producers*).

Providing *producers* with the tools for specifying how their preferred applications should be augmented is a good solution to help them to augment the Web from mobile devices and, at the same time, to create better mobile Web experiences. This strategy is reasonable since many recent studies have demonstrated that there is a global tendency of end users meeting the concrete needs of domain-specific scenarios by creating their own applications by using End-User Development (EUD) tools [35; 39].

In this work, we extended the MoWA approach with the aim of creating a general purpose authoring tool, in such a way that end users can develop their own applications without the need of having programming skills or even having to write a single line of code. As the main contributions of this paper, we aim to:

---

[2] Geolocation Web API: https://developer.mozilla.org/en-US/docs/Web/API/Geolocation
[3] Youtube's Data API: https://developers.google.com/youtube/v3/docs/search/list
[4] Twitter's REST API: https://dev.twitter.com/rest/reference/get/geo/search
[5] Firefox for Android SDK: https://developer.mozilla.org/en-US/Add-ons/Firefox_for_Android
[6] Device orientation event: https://developer.mozilla.org/en-US/docs/Web/API/DeviceOrientationEvent
[7] As in this demonstration: https://webrtc.github.io/samples/src/content/getusermedia/volume/

- Analyse how to overcome the challenge of developing Mobile Web applications by using an EUD approach based on augmentation.
- Outline an approach in which *developers* create domain specific components called *builders*, which are composable constructs available through an authoring tool, with the aim of empowering end users (*producers*) with the capability of creating domain-specific applications.

The remainder of this paper is structured as follows. Section 2 presents some background in regard of the main topics faced in this approach: End-User Development, Mobile Applications, Web Augmentation and our previous approach, called MoWA. In Section 3, we present our contribution: an End-User Development approach for Mobile WA applications. Section 4 presents our supporting tool and a case study. Evaluation procedures and results are presented in Section 5, involving participants with diverse characteristics: education levels, fields of study, ages, genders, used to different mobile platforms, using mobile devices with different frequencies, and having diverse levels of expertise in the use of such technologies. Section 6 outlines the existing work in EUD concerning Web applications, Mobile Applications Development –from both, desktop and mobile platforms– and Mobile Web Applications Development. Finally, Section 7 draws the conclusions of this work.

## 2. Background

### 2.1. End-User Development

Some studies, like [35; 39], indicate that there is a strong tendency demonstrating that the end user is starting to create, modify or extend their own software artefacts by using programming environments that abstract, somehow, the complexity of the development. This tendency gave rise to what we actually call EUD [27; 29], and was motivated by the need of users to quickly build their own solutions to the needs they have in their daily lives or circumstantially. Different to "traditional" software engineering approaches, in EUD the same person plays the role of developer and end user; he is the one who knows his context and needs better than anybody, and that not necessarily has formal education in development processes. EUD comprises a set of methods, techniques and tools that empower the user with the capability of creating, modifying or extending software artefacts.

For achieving the aforementioned, EUD relies in some well-known programming techniques, like Programming By Example (a.k.a. by Demonstration) [24], Extended Annotation (a.k.a. extended parametrization) [29], Visual Programming [9] and Scripting Languages [29]. The first technique consists in recording the sequence of actions of an end user in a system, so the application is built on the specification of the user's demonstration, and it does not require the user to code. The generated application should reproduce the commands he executed, and also allow him to parameterize some data objects used in the demonstration. The main benefit is that the user is writing programs through the user interface he is already familiar with, but it limits the end user to use already existing functionality in the base system or to add control structures to the recorded program. The second one is about associating a new functionality to the annotations the user makes, for example, allowing the user to annotate the DOM elements of a Web Page and associating a specialized behaviour to the element, as in [19]. Visual Programming is another set of techniques, all of them intended to build a program by letting the user to combine visual constructs. Languages here have a visual counterpart, a representation of the available conceptual entities and operations –the programming constructs–. In contraposition, text-based languages cover just one dimension and requires the user to learn the construct names and a syntax. Although this last technique can be a bit confusing –because it is hard to imagine an end user writing code– there are already widely used tools that implement this technique, and which are considered traditional examples of EUD, such as applications created using formulas in spreadsheets.

EUD applications started spreading in the Web, where large number of users joined different online communities to create and share their applications through public repositories. For instance, Userscripts[8] and GreaseFork[9] allow users to share Greasemonkey[10] scripts that adapt and augment the Web. Even traditional applications, such desktop spreadsheets, began to be conceived as part of the global Web scenario with the use of online, shareable and multiple access applications, like Google Sheets[11]. In the mobile applications field, EUD raised, expanding its scope and representing new challenges of integration with the possibility of providing both, the development process and the resultant application, with features based in diverse context types, like positioning, orientation or noise perception level. Studies like (Li et al., 2013) have been performed and it demonstrated –despite specific issues concerning their respective work– the steady growth of this new tendency.

---

[8] Userscripts mirror: http://userscripts-mirror.org/
[9] GreaseFork: https://greasyfork.org/en
[10] Greasemonkey: http://www.greasespot.net/
[11] Google Sheets: https://docs.google.com/spreadsheets/

### 2.2. Mobile Web Applications

One of the main benefits brought by the development of Mobile Applications [20] was the possibility of providing applications with mobile features (e.g. location-based) to offer the users customized services according to their environment. For instance, Context Aware applications [36] constantly monitor the users' environment and adapt the behaviour of the application accordingly. Mobile Hypermedia applications [5] make use of the position of the users and the Points of Interest (PoI) for assisting them in their navigation through the real world. There are also Mobile Augmented Reality applications [26], that consider the user's position for computing the position of digital objects in the real environment and draw them into a virtual layer.

During a long time, this kind of applications were developed with native code or specialized intermediate frameworks, which allowed developers to create native or hybrid applications [10]. At that time, Mobile Web Applications had little significance because there were no mechanisms for direct access to the device's internal services. Then, some works started to develop strategies for allowing Web applications to access the context information. For example, in [18] the authors presented a browser that interpret applications enhanced with specific XML tags. Such tags requested specific context information, and the information communication was achieved through the implementation of RESTful web services or POST requests. In [23] the authors adapt existing Web applications according to the context of use –e.g. sensing light and noise levels, hearth rate, user movements– resolving the adaptation content at server-side and then delivering the modified version for user consumption.

Nevertheless, due to recent advances in mobile Web browsers, Web applications can use the devices sensors values for adapting their behaviour, and new and interesting functionalities can be created.

### 2.3. Web Augmentation

Web Augmentation [4; 17] is a technique for manipulating and enhancing existing Web pages with new features. This makes it possible to meet needs that were not contemplated when a Web application was implemented, either because a) there was no such need at that time and could not be anticipated, b) it was not detected at the requirements elicitation stage, or c) simply because it was intentionally ignored. Augmentation is suitable for third-party stakeholders with very particular interests that have not been implemented; augmentations might also convey useful information for the owners of Web applications, because it can be used to identify uncovered needs of customers. For example, in Figure 1 we can appreciate how Télam, a news' portal, can be augmented with the capability of looking definitions in an online encyclopaedia, when the user holds a word on the screen of his mobile.

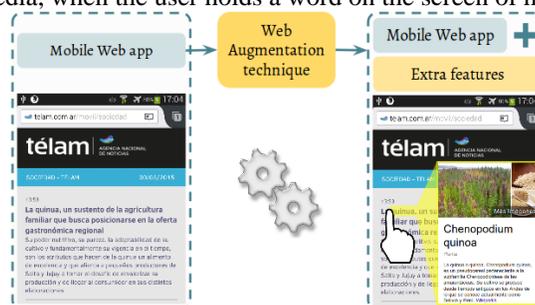

**Figure 1:** Web augmentation technique exemplification on mobile

There are several ways for implementing WA, but in general terms, we can classify them according to where the process takes place: at client-side or remotely (usually proxy-servers). In all cases, augmentations are materialized through the manipulation of Web Sites DOM, which is basically the UI that users perceive on client-side.

Web augmentation on client-side is appealing since different users can share or eventually improve the same *augmenter* on their own devices, without depending on third-party proxy servers such as transcodings [1; 17]. On client-side, there are several ways to address the deployment of such augmenters, but usually it implies installing some Web browser extension. These extensions may be found in the corresponding browsers markets (e.g. Firefox, Chrome, etc), and they are usually designed for augmenting a specific Web application (for instance, Amazon, Youtube, etc.). There are other kind of extensions which act as weaving engines enabling the execution of augmenters written in a Web browser agnostic way. Such augmenters are usually referred as userscripts. Most of these engines are available for all well-known Web browsers; they allow the reuse of the augmenters, something that is not possible with specific Web Browser extensions.

Without taking into account the deployment strategy, we can find several hundred thousand augmenters. The most used, such as *Magic actions for Youtube*[12] have far more than two million users, others such as *Flatbook*[13] (an extension for augmenting Facebook) near to one million, and others less known around one hundred thousand users, such as *Plus for Trello*[14] or *Koc Power Bot*[15]. There are even "official" extensions that provide better experience to customers, such as Amazon Assistant[16], with more than three million users. Besides that, all these repositories allow users to contribute and send feedback about the augmenters. The reader may see that for the most used ones, there are hundreds of comments suggesting changes or improvements. A deeper survey about exiting artefacts for the augmented Web can be found in (Diaz et al., 2015).

### 2.4. MoWA

In a previous work [3] we have presented an approach for augmenting exiting Web Applications, that originally do not contemplate mobile features, with these kinds of behaviour. The approach, called Mobile Web Augmentation (MoWA), comprises an augmentation framework and a weaver for running client-side scripts (MoWA applications). Developers instantiate the framework and create, as a result, a MoWA application. As shown in Figure 2, applications are installed into the MoWA weaver, a mobile browser extension, and their underlying augmentations are triggered either because the end user manually navigated to a target page, or because it was loaded in response to a change in the context (e.g. the user position). In both cases, the weaver runs the corresponding MoWA applications, which –in general terms– modify the UI adding extra components. Such components reside in an augmentation layer and they are adaptive to the user's context; this means that style, content or behaviour may be adapted when some context value change.

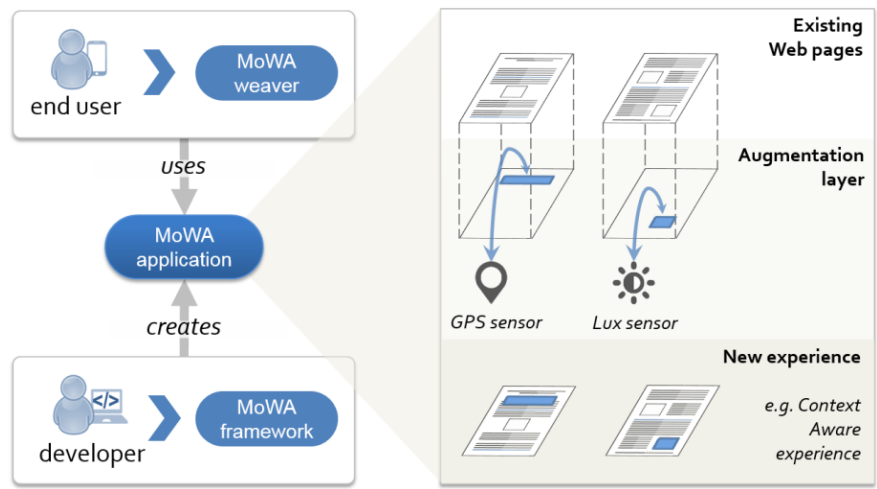

**Figure 2:** The MoWA approach

MoWA is based on the definition of *sensors* –observers– of *context types* [2], associating them with a set of augmentation components that will be placed over the content of a target Web page when a proper change in the observed context happens. For instance, consider a MoWA application that augments Youtube by adapting the application to changes in the noise level perceived by the phone; the volume of a video could be increased when it is being played in noisy environments, and decreased in the silent ones. Reusing content (from the target Web page to augment, and from some external ones) and accessing the mobile sensors is possible because the framework runs under the context of a mobile browser extension, which has access to the proper low level APIs.

---

[12] Magic actions for Youtube: https://chrome.google.com/webstore/detail/magic-actions-for-youtube/abjcfabbhafbcdfjoecdgepllmpfceif
[13] https://chrome.google.com/webstore/detail/flatbook/kadbillinepbjlgenaliokdhejdmmlgp
[14] https://chrome.google.com/webstore/detail/plus-for-trello-time-trac/gjjpophepkbhejnglcmkdnncmaanojkf
[15] https://greasyfork.org/es/scripts/892-koc-power-bot
[16] https://chrome.google.com/webstore/detail/amazon-assistant-for-chro/pbjikboenpfhbbejgkoklgkhjpfogcam

MoWA requires some basic knowledge of JavaScript to instantiate and combine the framework features, and it offers several hot spots that are easy to extend. Some of them address specific behaviour related to the supported context types. For instance, supporting orientation requires to implement a sensor but also a proper context type, in order to define some concrete context values of interest for the application. In this case, the context type is the *PointInSpace* class, which has one attribute for each axis in the tri-dimensional space, but other possible *ContextTypes* are: *Lux*, *Decibel*, *GeoLocation*, *QRBasedLocation*, *BatteryCharge*, *TelephonyStatus*, *UserProfile*, etc. Other hot spots deal with the augmentation aspects, enabling the incorporation of new augmenters, interpreting new kinds of context values, or supporting new dimensional spaces (e.g. an outdoors map, a 2D floor plan, a decibel scale, a brightness scale, etc.), domain functionalities, etc. Summarizing, MoWA empowers developers with a set of software features bringing together both, the mobile and the augmentation worlds.

Summarizing, the underlying concepts for a concrete MoWA application are:

- An existing *Web application* to be augmented or adapted with mobile features.
- The specification of the *context types* that will be used for the mobile experience. These types are based on a particular mobile device capability, such as GPS geo-localization or light sensors. Context values are made available to applications through software *sensors*, which are either part of the MoWA framework or can be defined by the developer. For instance, the location, as a context type, can be tracked for assisting the end user to traverse a tour's path.
- The association of sensors to a particular URL (among the ones already defined for a concrete application), to be opened and augmented when interesting events are notified by those sensors.
- A set of *augmenters* that specifically implement the augmentation layer considering the underlying application domain and the observed sensors.

While validating the approach [3], we gathered opinions and suggestions from participants; they emphasized the need of tools for automating some of the tasks they had to perform; for example, gathering the information required the definition of the points of interest and the positioning of markers in the map. Since such tasks implies a constant repetition of code and actions, we decided to create a visual tool such that it could be use not only by *developers*, but also by *producers*.

## 3. Our approach

In our research, we aim to empower end users with 1) the capability of augmenting existing Web sites with mobile features, 2) according to their own requirements, and 3) from their own mobile devices (although it may be done form desktop computers). It is worth mentioning at this point that there are some platforms allowing the authoring process from mobile environments [15; 31; 33; 38], but none of them conceives the creation or adaptation of Web applications; they depend, at some point, on a native or hybrid functionality. There are also applications augmenting Web sites from mobile devices [7; 23; 40], but they lack of EUD support. None of them perform the enhancement as part of the mobile browser, and their operation depend on a native component (e.g. for sensing and propagating the user's position). We present further details concerning existing works in Section 6.

Our challenge was not just about providing end users with the tools for creating Mobile Web applications using augmentation, but also about finding if end users were able to use such tools. In this sense, we are not speaking just about the usability aspects of the tool; we also wanted to assess if they were able to easily understand and apply the required concepts for creating their own experiences, and this includes being aware of the concepts listed in Section 2.4. In this direction, we developed a domain-specific authoring tool for creating Mobile Web applications, to enable end users to create their own experiences on-demand and in-situ (both, in the real and digital worlds).

Below, we present some possible scenarios to be faced with our approach, the supported user roles, and the main technical issues underlying our approach.

### 3.1. Motivating scenarios

We next present different scenarios designed to show both the potential and the flexibility of our approach. Although different scenarios may be combined, for the sake of clarity we separated them into two categories: one focusing in the use of context-awareness in general and another, more-specific one, in tours, by connecting diverse points of interest. We are aware that finer grained scenarios can be devised and tools that are more specific can be developed for this aim; we briefly comment this possibility in Section 7.

### 3.1.1 Context-Aware Scenarios

The idea of augmenting existing Web applications with context aware features is about enabling mechanisms for retrieving context values from the environment through the device sensors so the application can adapt its content and behaviour accordingly. The following scenarios are characterized, precisely, by using context information in their augmentations:

- Mobile Multimedia: hundreds of Web sites containing embed multimedia resources (such as YouTube/Vimeo videos) could be adapted in order to play those resources according to some context types, like light and noise levels.
- Yellow Pages Web sites: this kind of Web sites could be improved considering the current user's location. In this way, the results for a specific search may be enriched with a map showing how to reach some of the resulting places from the current location. The mobile experience could start a step before, considering to (auto) fill the search form considering the current user's city.
- Mobile Cinema portals: in many cities, the cinemas make it public their billboard in a common Web site (e.g. in La Plata, www.cinemacity.com.ar). In these sites, visitors may see the available movies and their functions. We can adapt portals like Cinema City into a full mobile Web application supporting different functionality:
    - Reordering movies according to the current time: the movie functions could be sorted to give a quick overview of the following functions.
    - Recommending the nearest cinema for a specific movie: considering the walking time for the movie that a user wants to see, the application may recommend the best cinema to go.
    - Adding a map showing the path to the cinema: when the user chooses a cinema, the application could add a map showing the path from the user location.

### 3.1.2 Tour-based scenarios

Tour guides represent a typical example of mobile applications; they provide people with assistance and information about a set of Points of Interest (PoI) spread in a city, museum, educational establishment, etc. They usually consider: 1) a set of PoIs, 2) a predefined path for navigating among these PoIs, and 3) a method for sensing the user's current position (QR codes, GPS, etc.). Below, we present some examples:

- Fixed indoor exhibitions in museums. Some museums use QR codes to identify each piece of their collection, and scanning them makes the mobile device to access to a digital counterpart of the object. For instance, consider the codes generated by QRpedia[17] that allow redirecting to concrete articles in Wikipedia. However, these digital counterparts of the pieces are not always linked, and the user could benefit from having assistance for touring the full or a concrete subset of the collection.
- Itinerant exhibitions, designed to be performed at the facilities of any building or open space, presenting collection focused in diverse kinds of artistic, historical, cultural manifestations that can be represented no matter the place. The physical points of interest in this kind of tours could be distributed, for example, in different areas of a school, and could be linked to a digital object, such as a Wikipedia article. Augmentations can be offered based on the information provided in the catalogue of the tour, and also by retrieving public comments to encourage discussion and the development of a critical attitude in the visitors.
- Temporary exhibitions, where a single art gallery is in continuous change, presenting different collections of artistic works –usually for periods of less than two months–. Augmentations can also be performed over Wikipedia articles (if the exhibition does not have its own Web site), adding quotations from the authors, comments by the visitors, and touring assistance along the way.
- Outdoor tours through the most significant heritage sites in a city, like the one presented by some city halls, allowing tourists to read about the best places for experience and appreciate the rich architectural history of the city. Augmentations here can be performed over each PoI Web page in the site, providing the touring assistance and also retrieving public review content through the some Trip Advisor Content API[18].

---

[17] QRpedia: http://qrpedia.org/
[18] TripAdvisor content API: https://developer-tripadvisor.com/content-api/

### 3.2. Targeted user roles

Before deepening into details, it is worth to make a clear distinction between two end-user roles that are considered by our approach. In this research we distinguish between:

- Producers; who create their own applications by using an EUD programming environment.
- Consumers; the ones who install and execute applications (created by a developer or a producer).

In order to understand which are the requirements for being a producer, we have surveyed the literature to identify common learning barriers on end-user programming systems. According to [28] there are six learning barriers (design, selection, coordination, use, understanding and information) in the context of end-user programming systems. In this regard, and following some authors recommendations and findings [16; 29; 33; 34], we defined a producer's live-programming environment, mainly based on pre-built widgets and forms. In this way, from our point of view, producers require to know what kind of mobile experience they want to have on a target Web site, and which context information (and the associated sensors) they want to use. They need to be capable of filling forms, placing widgets by drag and drop and, in order to let producers to *understand* their creation, they should be able to manage the preview of the mobile augmentations in any moment and checking that the expected behaviour is working as expected.

### 3.3. Mobile Web Augmentation by end users

The main idea behind our approach comprises, at least, a producer creating a Mobile Web application with a domain-specific authoring tool and a consumer using such application through a mobile browser weaver. There are other possibilities such as developers extending or instantiating the framework and improving the end-user tool, but we will concentrate on the former for the sake of clarity and conciseness.

We provide three tools to support this process: 1) MoWA authoring, an extension to Firefox for Android, supporting the creation of Mobile Web Applications at client-side; 2) MoWA crowd, an online platform supporting crowdsourcing and sharing services for MoWA artefacts; 3) MoWA weaver, another extension to Firefox for Android supporting the management and execution of MoWA applications.

As shown in Figure 3, a producer can interact with MoWA authoring, basically, for creating Mobile Web experiences from existing Web applications. He can create such applications based on his own requirements, but he can also do so upon the request of consumers, materialized in an app request posted in MoWA crowd. In both cases, creating an application may involve the extension of an existing one; to do so, the user must select an existing application, either in the local storage or in the remote repository. For the second option, he should search and download one of his applications in the repository, or a public one created by another user. The user may follow a series of steps that we explain later in Section 4.2 and, at the end of the process, it involves automatically saving a copy of the created application in the local storage of the browser extension, so it can be further edited or just executed in the producer's EUD environment. Finally, the producer has the chance of sharing his application in MoWA crowd. He can do so with the aim of sharing a new user experience, or for completing the app request process started in the MoWA crowd context.

A consumer can install MoWA weaver in his mobile browser, and start experiencing the augmented Mobile Web. To do so, he may execute MoWA applications that may be retrieved from the MoWA crowd repository or the ones he already has in his local storage (in the case he already has imported any or if the same user plays both roles). If the consumer cannot find an application facing his requirements in the public repository, he can start a new app request in MoWA crowd. Finally, and for the sake of space, we just mention that a consumer can also manage their local applications. E.g., he can edit or delete them, change the order of execution in relation with other applications, etc.

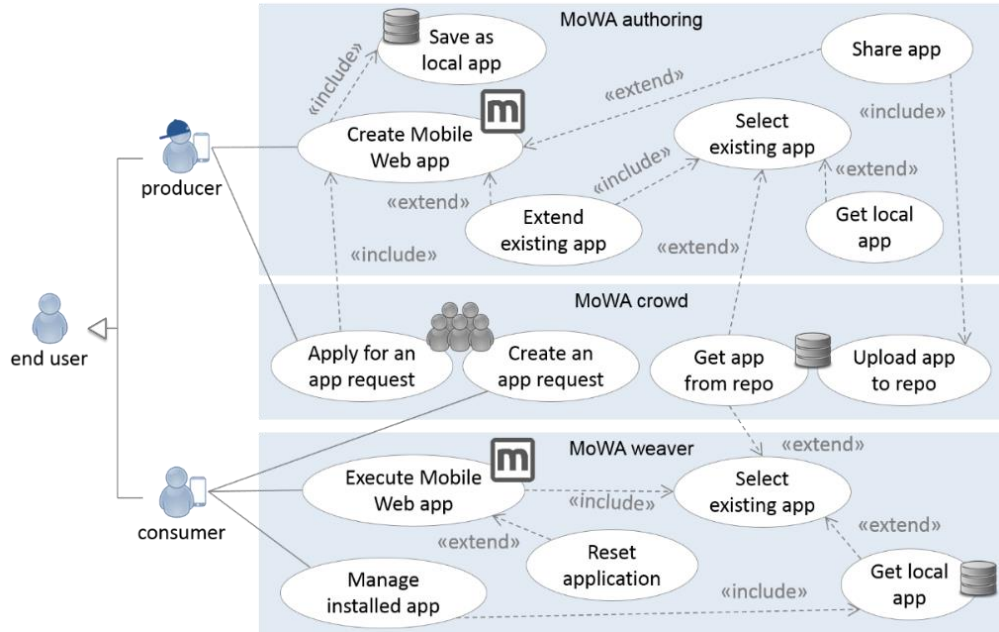

**Figure 3:** Steps in the authoring process

## 4. Supporting the authoring process

In this section, we present the architecture of the tool supporting our approach, the authoring process and the involved steps, and finally, we use our authoring tool for solving and presenting a case study, by following each step of the mentioned process.

### 4.1. The underlying architecture

The software architecture supporting the approach is presented in Figure 4. Two mobile tools are shown at both sides of the figure; they are in charge of enabling the authoring process and executing the resultant applications. There are also server-side components supporting application sharing and some crowdsourcing services, so consumers can demand for new solutions and producers and developers can meet such need.

Both client applications are Firefox for Android extensions, running on version 41 of such mobile Web browser. However, they have successfully run in precedent versions, as discussed in Section 5.2. As shown in Figure 4, you can see that both applications share a common base. First, the same mobile Web browser, that allow them to access the underlying capabilities of the device; for instance, it exposes multiple APIs for accessing the battery, the camera, the geolocation, the orientation, etc. This means that developers can access values of the user's context and take advantage of them in a direct way. E.g. by using the navigator.geolocation.getCurrentPosition[19], but also by using the APIS for processing data and track other context types. Example of this is accessing the camera, asking the user to point a QR code, taking it a picture and analysing it for decoding a QR code, which contains information related to the user's position. Second, they share the same base framework for instantiating, extending and executing applications: the MoWA framework. Finally, they also share some generic components for managing the user's installed applications in a local storage, but also for connecting to the server and upload, update or download applications.

At the right of Figure 4, you can see the customer's device, which has installed the MoWA weaver extension. He can import already defined applications from our repository, and use it for visiting existing Web sites with augmented features (e.g. he can download applications solving the aforementioned scenarios in Section 3.1). Such applications can be created in two ways: by client-side scripting (for developers) or by using the authoring tool (for producers). The first ones are downloaded from the repository as JavaScript files, and our weaver has a specialized interpreter in charge of loading the required classes for each application and instantiating them in the context of a set of concrete Web pages [3]. In the second case, another engine interprets authored applications, the ones created with MoWA authoring, that are materialized as an XML file, and specified according to the XSLTForms data model. This engine is in charge of interpreting such specification, instantiating the proper classes with the values in the data model, and then also cloning such objects in the context of the original Web page to augment.

---

[19] Firefox for Android's geolocation API: https://developer.mozilla.org/en-US/docs/Web/API/Geolocation/Using_geolocation

Then, in the middle, we have MoWA crowd, an online platform supporting crowdsourcing tasks management and also a repository of MoWA artefacts, among them, a collection of applications ready to be installed and used by *consumers*. Concerning the source of such applications, they can be both: the ones created by developers, as discussed in [3], or the ones created by producers with the MoWA authoring tool.

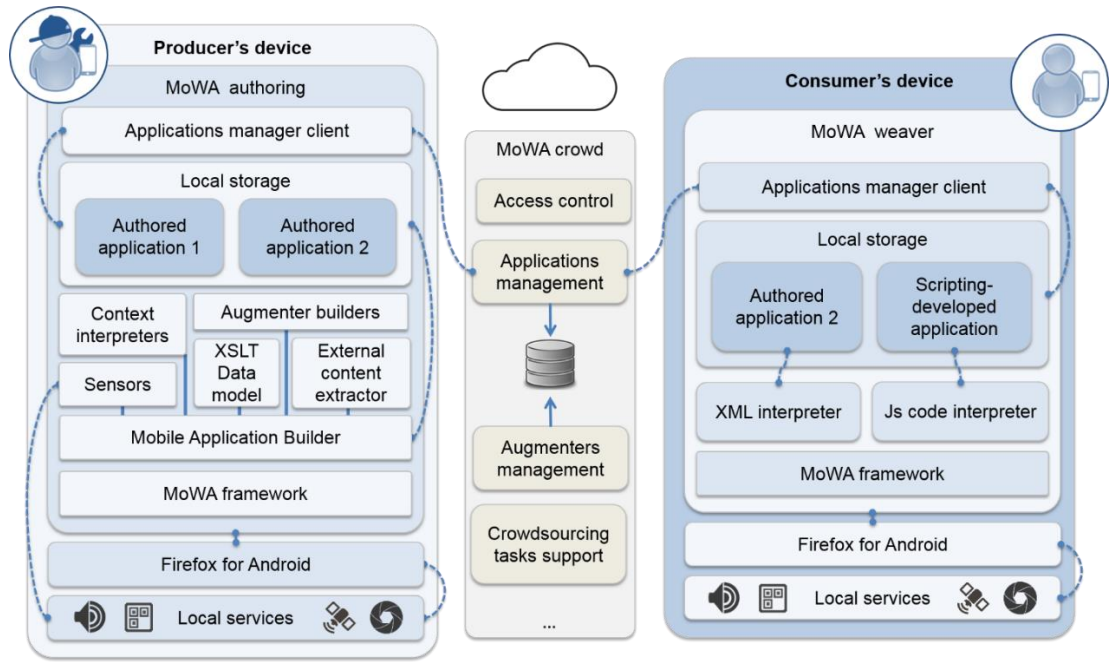

**Figure 4:** High level representation of the MoWA platform architecture

At left of the figure, the MoWA authoring tool architecture is presented. It is installed as an extension of the mobile browser on the producer's device. It comprises a set of specialized classes that decorate the base classes composing a MoWA application and extend their functionalities with the ability of asking the end users for the required parameters through a form-based assisted process, in order to instantiate such classes with the required values. For every configurable artefact supported in the authoring process, we have a decorator [22] wrapping it; we call them builders, and they assist the user in the authoring process by helping to configure specific instances to run properly.

There is a main builder, the one in charge of orchestrating the full authoring process: the Mobile Application Builder. It knows all the builders in charge of the application's construction process and their dependency to execute the application while it is being authored. The order matters; consider an augmenter dependent of the user's position. It can not be properly set up if, at least, a Web page has not been defined and if a sensor has not been selected. Therefore, it is important not to allow customers classes to access those subsystems directly. This way, the Mobile Application Builder also plays the role of a façade, being on charge of delegating the configuration of the application's components to other builders in the subsystem, in a specific order.

Builders might also depend on some values of the user's context. For example, consider a user walking a city and building a tour; he may want to retrieve his current position for creating a marker in a map, or detecting noisy areas in the city for defining an augmenter that increases the volume of a video. If the user is building and indoor tour based on existing QR codes in the building, he needs the tool to be capable of decoding those QR codes in order to associate the data to a physical position. And if the user selects a context-dependent augmenter for his application, it also needs to be subscribed, somehow, to the changes in the context. This depicts how the in-situ authoring process also depends on the MoWA sensors to be notified when their values change.

Our authoring tool provides end user with the capability of reusing existing content, but not just taking it as a base for the augmentations; it also allows him to extract it from external Web pages, usually third party ones, to be injected into a new context. For example, he can take the actors' profile at IMDB as target Web pages, and augment them with a carousel of related trailers from Youtube videos when the device is in landscape orientation. The content extraction is responsibility of a common component, available for every builder, named External Content Extractor. Such component is instantiated in the privileged context of the browser extension, so it makes it possible to append extra behaviour in any Web page, enabling user interactions for selecting their DOM elements of interest. This also makes it feasible the manipulation of every DOM element to obtain its positioning data (in the DOM, e.g. the XPath) and dynamically consume their content from external contexts (other Web pages that do not share the same domain name).

In order to persist the data entered by the user and keep it during different browsing sessions, we decided to use the data model supported by the XSLTForms engine [14]. XSLTForms it is a client-side implementation of XForms, whose benefits were extended by adding sub-forms management that allows us to easily support the authoring process in a wizard mode. The input values in XSLTForms are automatically binded to a data model that, at the end of the process, we can export and use as the specification of a MoWA authored application.

Back to the builders, they must be a subclass of *ConfigurableComponent*; it states that every builder should be capable of –at least– carrying out an authoring process, persisting and validating the data entered by the end user. Nevertheless, there is a need for a common interface between the builders (e.g. *MobileAppBuilder*) and the artefact to be authored (e.g. *MobileApp*), and that justifies the existence of components (e.g. *MobileAppCom*). In our model, components play the role of homonym name in the Decorator pattern, and each one of them inherits from *ConfigurableComponent*. All the components in Figure 6 are the darker classes with a name ending in «*Com*», and all the builders are the lighter ones ending in «*Builder*».

Each builder is responsible for defining the backbone of the part of the authoring process it is in charge. For example, the *MobileAppBuilder* implements the main algorithm process that is further explained in Section 4.2. At this point, it is enough to understand that the *MobileAppBuilder* is the most important builder; it is in charge of orchestrating the full process, and delegating tasks to more specialized builders, like the ones in charge of configuring the context values (*DimensionalSpaceBuilder*) or the augmenter layers (*ALayerBuilder*) and their augmenters (*AugmenterBuilder*). At the beginning, this builder asks for the basic information of the application (e.g. name) and then, in the following step it presents the end user with a series of context types supported by the *Sensors*. Consider a scenario where *location* was selected as one of the target context types and the *GPSSensor* as the only sensor to use. The application can delegate part of the authoring process to any *DimensionalSpaceBuilder* subclass (e.g. *2DMapBuilder*), in order to set the context values of interest for an *AugmentationLayer* of the *MobileApp*. So, following the example, the user is presented with a map, so he can define on it some markers matching a concrete coordinate. Another part of the process is delegated to the *ALayerbuilder* and it does the same with its *AugmenterBuilders*. When configuring the layer and augmenters, as the previous steps already gathered all the required data, it is possible to run the application and preview the augmenters while configuring them. The last step requested by the *MobileAppBuilder* requires the definition of a context rule that is materialized as the subscription of the authored layer to a concrete sensor. In each step, builders should persist the data entered by the end user, so at the end of the process it is not just possible to run the application but to export the full data model.

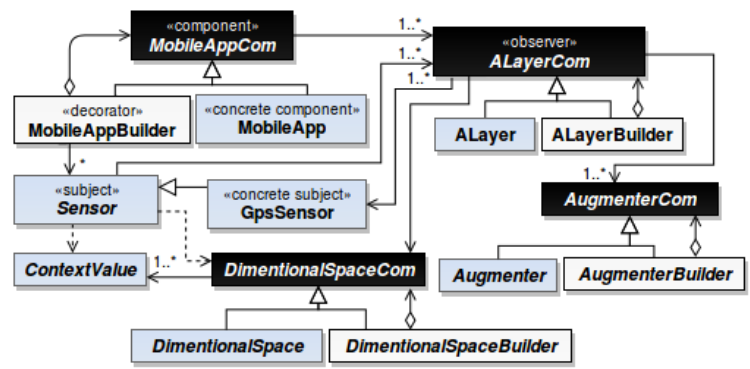

**Figure 5:** Decorating the MoWA framework classes with authoring features

As enabling user interaction with the authored artefacts may entail errors at execution time, MoWA Authoring components also support mechanisms for displaying extra messages to the end user in case an authored object was misconfigured. For example, if a map contains a wrong coordinate value, if the link of the image selected as a 2D floor plane is broken, or if an augmenter lacks of a required parameter value. Achieving this requires that any *ConfigurableComponent* may be capable of checking the arguments it is receiving and be capable of displaying proper and useful messages to the *producer*. Such class has an abstract method intended to that end: the *checkInputParameters*, which is executed before saving the configuration of a component (while in authoring mode), and before execution time (while in authoring or regular execution mode with the weaver).

Broadly speaking, we introduced new kinds of interactions to the framework components, like the ones required for managing the dimensional spaces. For instance, the case of 2D floor plans we used Leaflet Maps[20] (in conjunction with OpenStreetMap[21]), which provides a complete API for interacting with maps from both, mobile and desktop environments in a light way. Thus, we support two interaction modalities with maps: data visualization and authoring mode. In the first case, augmenters can use it for displaying the positions of the user and some PoIs, while in the second case it allows end users –through a concrete builder– to create, move, delete and connect markers in a map, and also to configure the map's zoom through touch events.

It is worth mentioning that there are no restrictions about the functionality for extending the builders. It is possible to contemplate manual mechanisms, where the user explicitly inputs the information, but also more sophisticated ones that autocompletes such data. For instance, we added a mechanism for facilitating the *in-situ* development modality to enable the possibility to add a marker at the user's current position. When the user needs to configure some Location values of interest for the app, he can: 1) insert and position it by tapping a special button with a positioning icon in the interface, or 2) hold on the map to insert the marker and then drag it to his desired position. The addition of such functionality also facilitates the validation of functional requirements, since it is the same person who sets out the requirements for the application the one that builds it under the same context in which it will be used.

Finally, we also adopted a definition of language bundles for each building artefact for internationalization purposes, so the MoWA engine is able to provide the authoring experience according to the user preferences or the browser's language. We provide language bundles of Spanish, French and English, and this allowed us to invite a broad spectrum of participants to our experiment; in fact, we conducted the experiment in the facilities of our laboratory in Argentina, and we had two participants in the experiment whose mother tongue was not Spanish. Nevertheless, they opted to create their application in Spanish.

### 4.2. The overall control flow

For supporting the authoring process of a mobile Web application (the create Mobile Web app use case in Figure 6) we started by defining it as a series of stages, represented as activities in Figure 6.

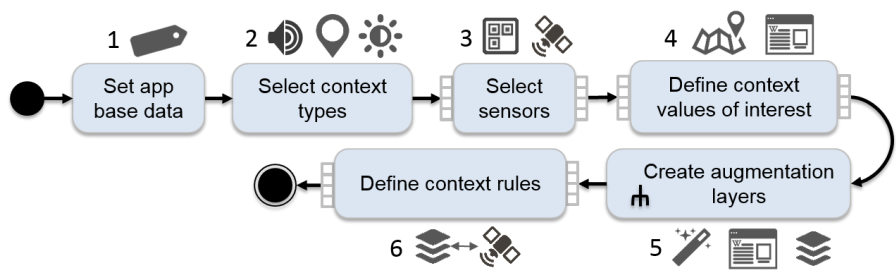

**Figure 6:** The supported process by the MoWA authoring tool

These stages are:

1. Setting the application base data. Here, the tool asks the user the base information for the application, like a name, a namespace, a filename, etc. The builder is in charge of getting the values that the end user should provide; in this case, just the application name. The rest of the required data is transparent for the end user.

---

[20] Leaflet Maps: http://leafletjs.com/
[21] OpenStreetMap: https://www.openstreetmap.org

2. Selecting context type(s). Here, the tool presents the end user with some context types, so he can choose among them and use these in his application. Examples of context types are the user's location, the device's orientation, the noise level, the light level or the time.

3. Select context sensor(s). Then, for each selected context type, the user is offered with a set of available sensors for listening for their changes. For example, both a GPS sensor and a QR sensor, could be on charge of sensing the user's location; a Lux sensor can notify about changes in the level of light perceived by the mobile device; a dB sensor can track changes in the noise level in the ambient.

4. Define context values of interest for the application. Every sensor notifies changes of a context type to the subscribed augmenter layers (that will be defined in the next stage), but in order to support such layers to use the sensed context values, it is required that the application knows which values are representative for her purposes. For example, as the end user is building a pure mobile application, the application needs to essentially know a set of locations for triggering the augmentations. Such locations are represented as Points of Interest and they have some optional properties that can be specified; for example, external content related to every point of interest or the specification of a navigability order through the set of PoIs. In the same way, an application subscribed to a Light sensor needs to know what are the bounding values that represent a significant change in the light level, and optionally it could be defined some associated data to every light level, as a default description of it. Therefore, every application is capable of showing a set of configurations according to the selected sensors.

5. Create augmentation layers. At this stage in the process, our tool asks the user to define a layer, and a set of configured augmenters for each Web page to augment. Concerning the first issue, there are two options for this purpose: the first one lets the end user to define a pattern that will be evaluated against the current URL in his browser, and the second one let him to select a concrete URL to open when a sensor notifies a layer to be executed. Concerning the augmenters, we provide the producer with a set of those artefacts according to the sensors he has chosen; we suggest them based on a simple tagging mechanism, defined as metadata in the augmenter's class file. Augmenters are defined in the context of a layer, and the producer can add as many as he wants. Each augmenter needs some input values to be properly executed. We provide the end user with three alternatives for defining such parameters' values: 1) he can reuse the defined data related to a concrete context value –e.g. the PoI name– by accessing the data model and selecting a property; 2) he can manually input such data in the form; or 3) he can use an assistant for retrieving external content.

6. Define context rules. Creating context rules is a transparent process for the end user, who simply must set the augmentation layers he has created as concrete observers of –at least– one of the selected sensors. The user is presented with the list of selected sensors and a set of augmentation layers, so he needs to specify which layers will be executed when a change in a concrete sensor happens. For instance, if he is using a GPS location and a Light level sensor, and he wants to execute different augmentations for each of them, he should create two augmentation layers with the desired augmenters for each of them, and finally match every layer with each sensor. In regard of the context rule composition, the event is represented by the sensor changes; the condition is the comparison of the sensed context value against the ones defined in stage 4; the action is the execution of a concrete augmentation layer.

### 4.3. The tool through a case study

Below, we present a case study matching every step of the aforementioned process in the domain of a Mobile Hypermedia tour in a museum; the process is similar in other domains and fields. We chose the hypermedia tour domain since the level of complexity required to build applications in this domain is much higher than the rest of the supported applications (e.g. those indicated in Section 3.1). In this sense, the end user must be able not only to configure sensors and augmenters to provide a Web application with context-based features, but also to configure navigational features at two levels; digitally and physically. This requires the end user to set up a space of representation, some points of interest with related information, and the walking links[22] between them.

---

[22] A walking link is supported by a regular anchor, but triggering it expresses the user intention to walk to a target location in the real world; it involves the physical movement of the user, as explained in [25].

The chosen scenario for this section was also used for the experiment set-up (See Section 5.3). At this point, it is only necessary to say that the museum has a collection of pieces that can be visited in a certain order, and each of them is associated with a QR code that redirects to a Wikipedia page. This museum also has its own Web site that provides information about the tour and each of its pieces –or PoIs–, but this information is not associated with the physical objects; the only digital counterpart of such physical objects are the Wikipedia articles.

In this context, a museum guide may want to create an application that augments the Wikipedia articles with the information in the museum's official Web site. To do so, he can carry out the authoring process, which contemplates the following stages:

1. Setting the application base data. As shown in Figure 7, once the authoring mode is enabled ($S_1$) and the producer chooses to create a new application ($S_2$), he should set a name for the application ($S_3$).

2. Selecting context type(s). Then, the producer should choose among the available context types. In this case, he has chosen just the location ($S_4$).

3. Select context sensor(s). According to the selection he did, he is offered in ($S_5$) with the configuration of a single context type, this is, the location. Then, in (S6) he must choose among the related sensors observing changes in location: GPS and QR-based sensors.

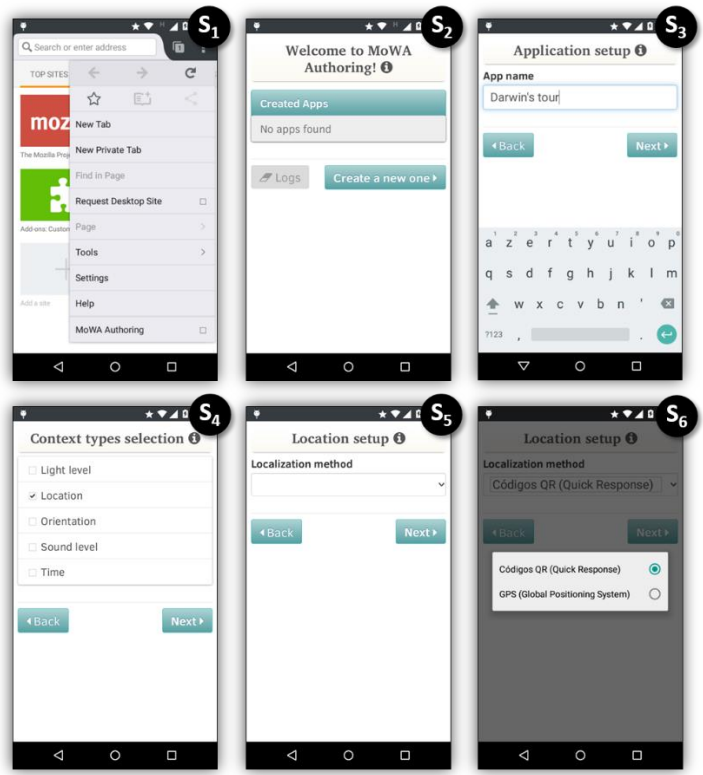

**Figure 7:** Choosing the concrete sensor

4. Define context values of interest for the application. In this case, as he has selected a QR-based location sensor, he should define locations by creating markers in a map and associating each of them with the data contained in a QR code. As we can see in Figure 8, the first step ($S_1$) requires the producer to configure a "dimensional space" for defining some locations as the context values of interest. The producer chooses "2D floor plan", that requires some extra configuration shown in $S_2$. Once an image has been chosen, it is also needed that the producer defines the PoIs, as explained in $S_3$. Such definition also carries some parameters to be configured, as seen in $S_4$; scanning a QR code, as in $S_5$ and then some external content can be extracted and defined as properties of the PoI. We can see in $S_{6,7,8}$ there are two properties: "poi-desc" and "poi-pic"; they will be used for augmentation in future steps. Once these parameters are defined, the producer is back to the same screen ($S_3$), but this time visualizing a marker on the map. If he holds it, he can access to a context menu ($S_9$) that allow him to go back to edit and connect the selected PoI, as finally shown in $S_{10}$.

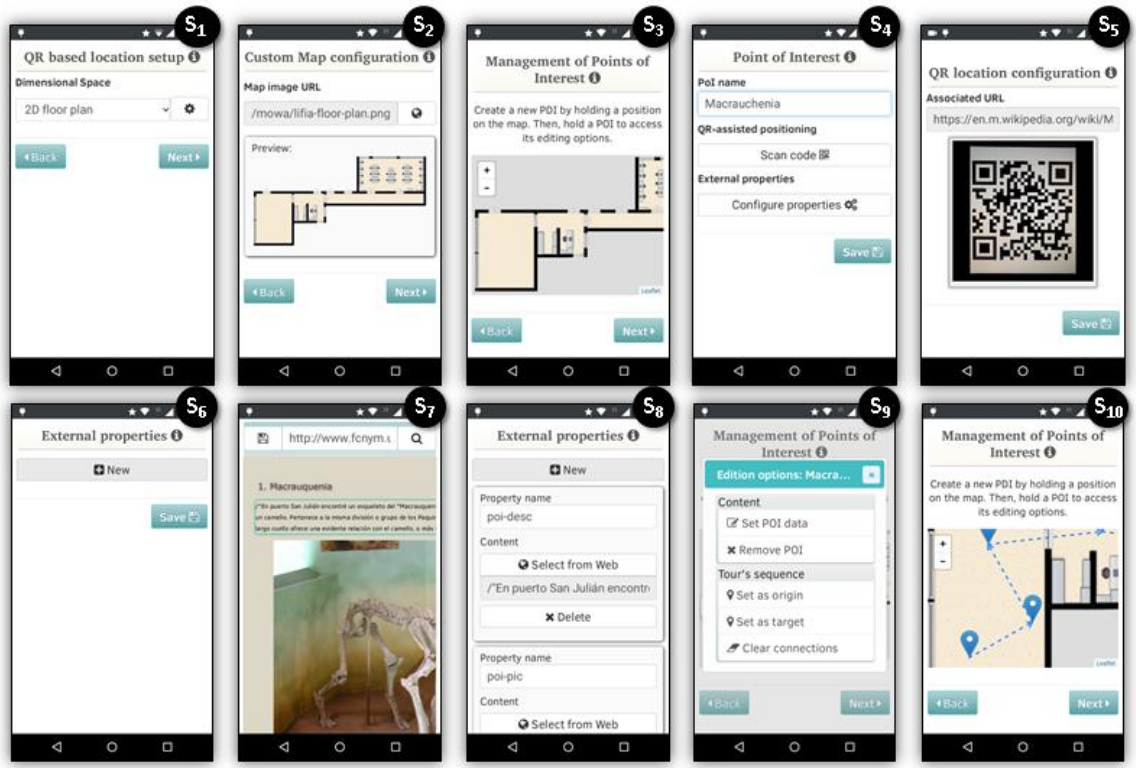

**Figure 8:** Configuring the concrete required parameters of the application

5. Create augmentation layer(s). As shown in Figure 9, the producer needs to create at least one layer ($S_1$) for each PoI. Then, he must define a target Web page(s) to augment ($S_2$) for which purpose he has chosen to define a concrete URL. So, in $S_3$, he should provide the URL to augment; he can do it by navigating the Web or selecting an already defined value in the application (e.g. the data decoded from the QR code). Then, in $S_4$, the user can add multiple augmenters on each layer for being executed together. For doing that, he needs to navigate the carousel at the bottom of the window and insert an augmenter by clicking on the proper item; this will insert a new floating augmenter-thumbnail at the middle-center of the screen, which should be positioned in the DOM by drag and drop as in $S_{4,5}$. Then, he must configure the augmenter's parameters shown in $S_{6,7}$. He can write the input or he can link the configurable parameter with the application given name, as in $S_8$. He is taken back to the Web page to augment and he can preview the configured augmenter, as in $S_8$. Once he creates as many augmenters as he wants, he can preview all the augmenters together and appreciate the layer's result, as in $S_n$.

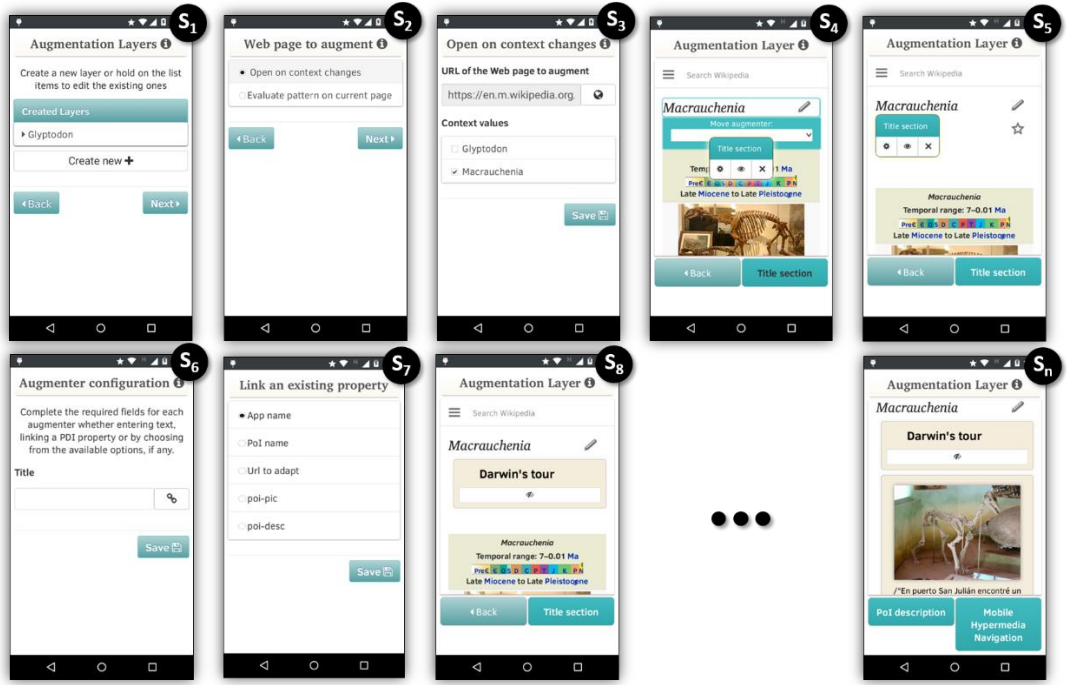

**Figure 9:** Augmentation layers management

6. Define context rules. In this case, defining a rule it is a matter of matching sensors with augmentation layers. The event is the QR-based sensor change; the condition is comparing the sensed location value against the context values defined in stage 4; the action is the execution of an augmentation layer (which may involve previously loading of a concrete Web page, as explained in stage 4). As in this case there is defined just a single sensor, you can see in Figure 10 that layers are automatically matched to such sensor ($S_1$) and the user just need to generate and export the application ($S_2$).

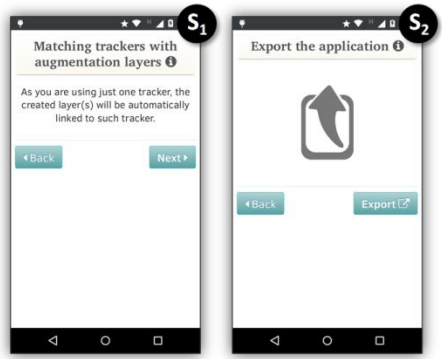

**Figure 10:** Defining augmentation layers triggered by a single sensor

As result of the process, a MoWA application specification is obtained, and the *producer* can share it with *consumers*. The MoWA weaver could be used for running such applications later. Let's say that the producer shares the script with Julio, his cousin, who is visiting the Museum for the first time. Julio installs the authored application by using MoWA weaver and, once in the Museum, he scans the first QR code he finds with the QR codes scanner provided by MoWA Weaver. Such action will load a Wikipedia article in the browser and, if the code is part of the ones contemplated by the application, the Wikipedia article will be augmented with the extra information taken from the Museum's official Web site. But such information will be adapted according to the context values. For example, if the visited piece is the one defined as the first in the tour, Julio will be presented with information about it and the path he should walk to arrive to the next PoI; otherwise, the augmentation will indicate him how to walk from his current position to the first PoI. The same will apply to the following PoIs; the tour will assist Julio with information about each piece and the tour's order.

## 5. Evaluation

Due to the complexity of our approach, the issues that can be taken into account at evaluation time are extensive. In this work we covered several domains, integrated diverse technologies and promoted the user-roles cooperation, therefore, we can conceive the idea of such tools being target of not one but several validation experiments. Nevertheless, as the main challenge of this work is related to the EUD process itself, we have focused the evaluation on the producers' activity.

We conducted the experiment in the context of a Mobile Hypermedia application, oriented to the tourism domain. We took MoWA authoring as the *experimental unit* for the *experiment design*, and the tool was used by end users for solving a concrete scenario problem, proving the feasibility of our approach.

In this light, we wanted to prove the following hypothesis:

> *Any end user using MoWA Authoring is capable of building Mobile Web Augmentation applications with a success rate of, at least, 80%.*

We formulated the null hypothesis ($H_0$) stating that end users can create Mobile Web Applications with a successful rate of 0.8, against an alternative hypothesis $H_1$ where such creation value is higher than such value. In this sense, we specified $H_0 = \sigma_1 = 0.8$, against the alternative $H_1 = \sigma_1 > 0.8$, where $\sigma_1$ represents the mean successful rate for applications creation.

### 5.1. Participants

We conducted the experiment with 21 experimental subjects with a very wide range of demographic characteristics. Due to the large number of participants in the session, we chosen surveys as collecting data mechanism. We found that 71.43% of the members were males, against a 28.57% of female ones. Their ages range was homogeneous, ranging from 22 to 39 years old, with an average of 26.86.

An interesting factor was the cultural diversity of participants: 66,67% of them were Argentinean citizens, but there were also participants with different nationalities: 14,29% of Colombians, 9,52% of Frenchs, 4,76 of Venezuelans and 4,76% of Peruvians. They were mainly ongoing students of different degree careers, representing the 52.38% of the population, preceded by a group of the same education level but different status: 28.58% of them already completed their studies. At equal frequency, with 4.76%, were other 4 groups representing: completed high school level, uncompleted degree career, completed post-degree career, and post-degree studies in course.

As we focused on real end users with a wide range of interests, we searched for volunteers in various contexts, involving people coming from different fields of study: most of them were from hard sciences, with a 52.38%, followed by a 19.06% from Social Sciences, 9.52% from Arts, 9.52% from Natural/Life Sciences and 4.76% from Economics. The remaining 4.76% corresponds to participants who did not follow specialized studies.

Concerning the technological expertise of the participants, we focused in three aspects that allowed us to infer their level of technological background and accordingly classify them to observe if there is any influence in their performance. Such aspects were related to their programming experience, technological know-how and the user perceived complexity in a series of related tasks related to our tool. As shown in Table 1, according to these three factors, we categorized them as novice users (N), regular users (R) and expert users (X).

Regarding the first aspect, we just ask them about their programming skills experience. A 47.62% never experienced programming, 42.86% of them programs frequently and a 9.52% occasionally programmed. We considered them as novice users, expert users and regular users, respectively.

In order to analyse the second factor, we presented them a set of questions based on how frequently they perform certain tasks. We asked about the usage of 1) smartphones, 2) the Android platform, 3) Web browsers in daily life activities, 4) Firefox for Android, 5) browsers' extensions, 6) Firefox for Android extensions, 7) location aware apps, 8) guided tours apps, 9) Web Augmentation apps 10) authoring tools, 11) mobile Web forms, 12) QR codes scanning, 13) Web searching, 14) content selection and edition, 15) map markers management, 16) slice gesture for menu navigation and 17) drag gesture for moving objects. We provided 5 options in each case: daily, weekly or monthly usage, 'a few times' and 'never did it'. The first option was related to expert users, the second and third ones with regular users, and fourth and fifth with inexperienced users. This way, we assigned every answer with an expertise level, and took the higher occurrence percentage of them for representing the user expertise level. For the sake of simplicity, we show final results in third column of Table 1.

We were also interested about the user's perceived complexity of specific tasks related to our tool, not using the tool itself but of similar UI interactions. Such tasks were a subset of the previously presented, ranged from the one numbered 11$^{th}$ to the 17$^{th}$. This time, the possible values were: very easy, easy, normal, hard, very hard. The criteria was the same applied with the previous procedure, so we matched the first case with expert users, the second and third with regular users, and the remaining two with novice users. We show the final results of this process in the fourth column of Table 1.

Finally, we assigned the users a category by selecting the predominant in the three analysed factors. When the three options are present, we chose the intermediate value –the Regular User–. This way, we have a population conformed by 28.57% of novice users, 28.57% of regulars and 42.86% of expert ones.

| PS | Programming experience | Technological mastery | Complexity perception | Prevalent category |
|---|---|---|---|---|
| 1 | N | X | R | R |
| 2 | X | X | X | X |
| 3 | R | N | R | R |
| 4 | X | X | R | X |
| 5 | N | N | R | N |
| 6 | R | X | X | X |
| 7 | N | N | R | N |
| 8 | N | R | X | R |
| 9 | N | X | X | X |
| 10 | N | R | X | R |
| 11 | N | N | X | N |
| 12 | N | N | X | N |
| 13 | N | N | X | N |
| 14 | X | N | X | X |
| 15 | X | R | X | X |
| 16 | X | N | R | R |
| 17 | N | X | X | X |
| 18 | X | N | R | R |
| 19 | X | X | X | X |
| 20 | X | X | X | X |
| 21 | N | N | R | N |

**Table 1:** technological expertise of the participants

## 5.2. Platforms

The MoWA authoring tool is implemented as an extension of Firefox for Android, so it was a requisite to use any Android-based device capable of running such mobile Web browser. Users carried their phones and installed the required software. The concrete authoring tool was successfully executed in every single device, which ranged among 5 different benchmarks and a total of 7 models. In the 4.76% of the cases, participants used a Blu Studio 5.5 mobile device; 4.76% a Huawei G6 L33; 9.52% a LG Pro Light; 28.57% a Motorola Moto G; 9.52% a Motorola Moto E; 38.1% a Samsung Galaxy S3 I9300; and 4.76% a Samsung Galaxy S2.

In regard to the Android platform version, the mobile phones have installed 5 different versions. Android Jelly Bean was present supporting three API levels: a 4.76% of the total devices supported the API 16 (v.4.1.2) version; another 4.76% the API 17 (v.4.2.2); and 52.38% the API 18 (v.4.3). Remaining percentage corresponds equally, with 19.04%, to Android KitKat for API level 19 (v. 4.4.4) and Lollipop for API level 21 (v.5.0.2).

Our extension successfully run in 4 versions of the mobile browser: the 66.67% in version 38.0 –the stable one at that moment–, the 23.81% in 39.0 –Firefox Beta–, the 4.76% in 37.0 and another 4.76% in 41.0 –Nightly–.

## 5.3. Experiment setup

We asked users to create a museum tour guide app with our authoring tool. We simulated the *Museo de Ciencias Naturales de La Plata* in the facilities of our laboratory, by printing the same QRPedia codes found in the museum and placing them in four separate –but connected– spaces of our facilities, as seen in Figure 11. We decided to do it in several spaces, trying to simulate the museum environment in the best way, so that users have to walk over a large area and look for the pieces.

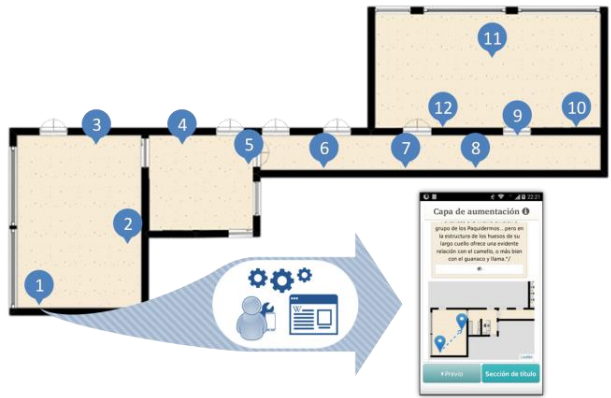

**Figure 11:** Physical distribution of PoIs, replicating the museum at our facilities

Participants weren't trained in the use of the authoring tool, so we explained our approach and some basis about the tool usage in a 30-minute talk before the experiment begins. They were introduced to Mobile Hypermedia applications, Web Augmentation and In-Situ development. After that, participants were asked to complete a survey for demographic data collection and user expertise analysis. Once every participant finished the survey, a scenario was explained to them:

*You are about to visit the Museo de Ciencias Naturales of the Universidad Nacional de La Plata, in order to perform a special tour, that presents a set of pieces related to the Darwin's findings and descriptions made during his Voyage of the Beagle. You knew about this tour thanks to the Museum's website, so you already read something about the pieces and also know they are presented in certain order.*

*When you arrive to the Museum, you observe that the pieces –not only the ones of the mentioned tour– have a QRpedia code. Then you realize that some information is not being presented to those visitors who didn't know about the Web site. You come up with the idea of combining the information submitted on the official Web site of the museum and the one presented in Wikipedia, and also incorporate certain functionality that assist the user in visiting the pieces in the right order. In order to do this, you need to create a Mobile Web Application that supports:*

- R1. *augmentation of the Web page related to each piece with new information extracted from the Museum's official site,*
- R2. *users assistance for showing the following piece to be visited, and*
- R3. *users assistance if they are in a wrong piece.*

Back to the current experiment set-up, once the statement was understood for the participants, they received printed instructions. They were explained they had to install a concrete MoWA authoring tool, whose creation process is assisted through a series of steps implemented as forms. We collected data of each individual experience by logging every user action, registering a copy of their final productions and asking them to complete some surveys. At the end of each step, the printed instructions requested the user to assign a level of difficulty of the performed tasks. At the end of the process, users were asked to complete another survey, this time about their perception in the usage of the tool.

### 5.4. Results

The collected data was analysed and then significance tests were applied in order to accept or refuse the concerning hypotheses. In Table 2, we present the results of the 7 steps performed by the 21 participants (*PS*). We were interested in knowing the completed tasks percentage for each participant, so we applied a concrete criterion for each requirement and obtained their completeness percentage.

| PS. | R1 | | | R2 & R3 | | | | SR |
|---|---|---|---|---|---|---|---|---|
| | a | b | % | c | d | e | % | |
| 1 | 0.98 | 0.50 | 0.74 | 0.50 | 1.00 | 1.00 | 0.83 | **0.80** |
| 2 | 1.00 | 1.00 | 1.00 | 1.00 | 1.00 | 1.00 | 1.00 | **1.00** |
| 3 | 0.71 | 0.71 | 0.71 | 0.92 | 1.00 | 1.00 | 0.97 | **0.89** |
| 4 | 1.00 | 1.00 | 1.00 | 1.00 | 1.00 | 0.80 | 0.93 | **0.96** |
| 5 | 0.33 | 0.17 | 0.25 | 0.17 | 0.33 | 0.30 | 0.27 | **0.26** |
| 6 | 1.00 | 0.00 | 0.50 | 0.00 | 1.00 | 1.00 | 0.67 | **0.61** |
| 7 | 0.58 | 0.58 | 0.58 | 1.00 | 1.00 | 1.00 | 1.00 | **0.86** |
| 8 | 0.92 | 0.92 | 0.92 | 0.92 | 0.92 | 0.50 | 0.78 | **0.83** |
| 9 | 1.00 | 1.00 | 1.00 | 1.00 | 1.00 | 0.70 | 0.90 | **0.93** |
| 10 | 0.67 | 0.33 | 0.50 | 0.33 | 0.67 | 0.50 | 0.50 | **0.50** |
| 11 | 1.00 | 1.00 | 1.00 | 1.00 | 1.00 | 1.00 | 1.00 | **1.00** |
| 12 | 0.54 | 0.54 | 0.54 | 1.00 | 1.00 | 1.00 | 1.00 | **0.85** |
| 13 | 1.00 | 1.00 | 1.00 | 1.00 | 1.00 | 1.00 | 1.00 | **1.00** |
| 14 | 1.00 | 0.83 | 0.92 | 0.83 | 1.00 | 1.00 | 0.94 | **0.93** |
| 15 | 1.00 | 1.00 | 1.00 | 1.00 | 1.00 | 0.60 | 0.87 | **0.91** |
| 16 | 1.00 | 1.00 | 1.00 | 1.00 | 1.00 | 0.70 | 0.90 | **0.93** |
| 17 | 0.85 | 0.88 | 0.87 | 0.92 | 0.88 | 0.70 | 0.83 | **0.84** |
| 18 | 1.00 | 1.00 | 1.00 | 1.00 | 1.00 | 1.00 | 1.00 | **1.00** |
| 19 | 1.00 | 1.00 | 1.00 | 1.00 | 1.00 | 1.00 | 1.00 | **1.00** |
| 20 | 1.00 | 0.54 | 0.77 | 1.00 | 1.00 | 1.00 | 1.00 | **0.92** |
| 21 | 1.00 | 1.00 | 1.00 | 1.00 | 1.00 | 1.00 | 1.00 | **1.00** |
| % | | | 0.82 | | | | 0.88 | **0.86** |

**Table 2:** percentage of completed tasks by participant

Criterion for R1 was checking if applications have defined (a) the 12 PoIs and (b) the PoI augmenter for the 12 PoIs' associated Web pages. In the first case, a percentage was obtained by analyzing whether each PoI have defined a name, a URL to augment and an external picture and description, both referenced from the Museum's official Web site. In the second case, we analyzed if the PoI augmenter has been rightly positioned and configured with the right external values. Each property was evaluated separately, and both made the final percentage value. When a property was partially configured, the property was marked as incomplete (0) and only right values with (1). Results shown in column b represent the average of both properties.

Criterion for R1 and R2 were the same, because both can be achieved by defining the Mobile Hypermedia Navigation augmenter, defining the 12 PoIs and connecting them. So, we checked these three issues and got and present them in columns c, d and e respectively. Values in column c represent a percentage, for each participant, of rightly positioned augmenters for the 12 PoIs; values in column d, the percentage of defined PoIs in relation to the 12 requested, and just considering the name and URL to augment as required attributes; column e, the percentage of connected PoIs in relation to 11 expected. According to this, successful rates (SR) were obtained as the average of the percentages of the three requirements, this is:

$$SR_i = \frac{\frac{a_i+b_i}{2} + 2 * \frac{c_i+d_i+e_i}{3}}{3}$$

Total result let us appreciate that in general 86% of the required tasks of the experiment could be completed by the users, with a standard deviation of 0.1879. But as the collected data do not correspond to a normal distribution we opted to use the mean rather than the median, and to perform a sample sign test to validate the hypothesis. We considered an α-level of 0.05 as our maximum acceptable level of risk for rejecting the null hypothesis, and a confidence level of 95%. We choose a hypothesized median of 0.8, that then we were also able to fit it with an optimal value of 0.84; among the 21 analysed observations, 5 of them are below that value, 1 is equal and 17 are above it. We performed the test and attained a significance level (p-value) of 0.0207. Therefore, as such value is less than α, the null hypothesis was rejected in favour of the alternative.

With regard to the times, the whole process took participants an average of 1:07h, ranging from 35minutes to 1:34h. We started logging when the "create application" option was triggered, and stopped when the application was exported. We logged every single interaction of the user and save it to a file that finally was uploaded, in conjunction with the application, to our server. These times allowed us to observe the real consumed time by the user for the whole process. The presented steps in Table 3 are: 1) creating a new application, 2) setting up the dimensional space, 3) defining the PoIs, 4) enhancing the related Web pages, 5) connecting the PoIs and 6) exporting the authored application. As we can see, steps 3 and 4 were the most demanding ones.

| PS | Step 1 | Step 2 | Step 3 | Step 4 | Step 5 | Step 6 | Total |
|---|---|---|---|---|---|---|---|
|  | *m:ss* | *m:ss* | *m:ss* | *m:ss* | *m:ss* | *m:ss* | *h:mm:ss* |
| 1 | 1:11 | 4:34 | 38:13 | 18:04 | 5:14 | 0:04 | 1:07:20 |
| 2 | 1:33 | 1:08 | 39:37 | 23:59 | 6:52 | 0:02 | 1:13:11 |
| 3 | 1:54 | 2:32 | 30:00 | 32:28 | 6:46 | 0:07 | 1:13:47 |
| 4 | 2:56 | 1:08 | 41:56 | 22:49 | 7:22 | 0:21 | 1:16:32 |
| 5 | 4:28 | 4:38 | 24:22 | 30:22 | 3:12 | 0:33 | 1:07:36 |
| 6 | 1:32 | 4:42 | 42:38 | 10:50 | 7:32 | 0:01 | 1:07:15 |
| 7 | 2:55 | 3:46 | 31:56 | 37:26 | 7:40 | 0:27 | 1:24:10 |
| 8 | 1:58 | 1:15 | 43:44 | 30:15 | 6:08 | 0:03 | 1:23:23 |
| 9 | 3:14 | 3:07 | 44:58 | 36:35 | 6:19 | 0:03 | 1:34:16 |
| 10 | 4:40 | 4:38 | 42:19 | 33:01 | 6:43 | 0:35 | 1:31:55 |
| 11 | 2:22 | 2:01 | 36:37 | 30:48 | 8:21 | 0:02 | 1:20:12 |
| 12 | 1:33 | 2:41 | 19:00 | 33:56 | 5:35 | 0:13 | 1:02:58 |
| 13 | 1:14 | 1:24 | 41:06 | 26:09 | 6:57 | 0:10 | 1:17:00 |
| 14 | 3:22 | 2:26 | 37:13 | 14:34 | 5:59 | 0:03 | 1:03:37 |
| 15 | 1:41 | 2:15 | 20:44 | 18:43 | 3:08 | 0:02 | 0:46:34 |
| 16 | 1:14 | 5:23 | 39:14 | 16:33 | 5:45 | 0:02 | 1:08:11 |
| 17 | 1:08 | 0:52 | 25:33 | 12:53 | 4:32 | 0:03 | 0:45:01 |
| 18 | 0:56 | 2:27 | 50:20 | 15:15 | 6:15 | 0:02 | 1:15:14 |
| 19 | 0:29 | 0:45 | 06:46 | 15:18 | 1:04 | 0:02 | 0:36:24 |
| 20 | 0:35 | 0:52 | 06:08 | 14:24 | 1:22 | 0:02 | 0:35:23 |
| 21 | 0:44 | 1:41 | 08:02 | 14:57 | 2:22 | 0:04 | 0:39:51 |
|  | 1:59 | 2:35 | 31:56 | 22:15 | 5:29 | 0:09 | **1:07:08** |

**Table 3:** percentage of completed tasks by participant

In relation to the population, we can say that women –the 28.57% of the population–, in average, completed the 89% of the requirements in 1h 1min, with an average difference between the dataset and the mean of 15.28%. Men did an average of 84% of the same tasks in 1h 9min but with a higher standard deviation, of 20.39%.

The 66.67% of the population were Argentineans, who completed in average the 84.47% of the tasks with a deviation of 21.23%. Colombians, a 14.29% of the participants, did the 85.56% of the tasks, with a deviation of 21.29%. Frenchs, who represent the 9.52% of the population, did the 90.44% of it, with a standard deviation of 2.67%. Finally, with just one representative for each nationality, Peru obtained a 100% and Venezuela an 82.67%.

Concerning people's formation, ongoing students of a degree career represent the 52.38%, and they achieved, in average, 90.82% of the experiment with a deviation of 14.74%. Degree graduates are the 28.58% of the population and they did the 88.64% of the tasks, with a deviation of 5.99%. The remaining population are four people, one for each of the following categories: high school graduate, with a 26.11% level of completeness; abandoned degree career, with 84.67%; post-degree graduates, with a 100%; and post-degree students with a 61.11%.

Most of the participants followed specialized studies, but one of them. This participant represents the 4.76% of the population and he did the 26.11% of the tasks. The ones belonging to hard sciences field (the 52.38%) did the 93.51% of the tasks with a standard deviation of 6.88%; the 19.06% coming from Social Sciences, did it the 89.81% with a deviation of 6.99%; the 9.52% from the artistic field, did the 88.86% with a deviation of 6.32%; and other 9.52% coming from Natural/Life Sciences completed the 80.56% of the tasks, with a deviation of 27.5%. Of course, the last two categories may not be as significant as the previous ones, because they have just two representatives each. Finally, there is just one participant from Economics field, who could do the 50% of the tasks.

Regarding the users' expertise, novice users (28.57%) completed the 82.8% with a standard deviation of 28.7%; regular users (28.57%) the 82.46% with a deviation of 17.45%; and expert users (42.86%) a 90.14% with a deviation of 11.86%.

From the technological point of view, three participants used one of the following mobile device models each: a Blu Studio 5.5, a Samsung Galaxy S2 and a Huawei G6 L33. They achieved the 95.56%, a 93.33% and finally a 100%, respectively. Users of LG Pro Light and Motorola Moto E were a 9.52% of the population each; in the first case, they achieved the 63.06% of the tasks with a standard deviation of 52.25%, in the second case, they did the 73.56% with a standard deviation of 17.6%. Users of Motorola Moto G represent the 28.57% and they completed the 82.9% with a standard deviation of 17.66%. Finally, participants using a Samsung Galaxy S3 I9300 (38.1%) completed the 92.91% of the tasks with a deviation of 6.66%.

Five versions of Android were used in the experiment. Two of them had a single representative; the API 16 (v.4.1.2) version and the API 17 (v.4.2.2). Participants using the first one completed the process in a 93.33%, while the second one in a 95.56%. Then, the API 18 (v.4.3) represent the 52.38% of the population that did an 88.13% with a deviation of 21.5%. The remaining percentage corresponds equally (19.04%) to API level 19 (v. 4.4.4) and API level 21 (v.5.0.2). In the first case, 72.63% with a deviation of 20.45%; in the second one, 88.5% with a deviation of 8.97%.

Our extension successfully run in 4 versions of the mobile browser, the –Firefox Beta–, the 4.76% in 37.0 and another 4.76% in 41.0 –Nightly–.

Finally, the extension run in different versions of the mobile browser. Two participants used versions 37.0 and 41.0 each, completing the 93.33% in both cases. The 66.67% of the population used version 38.0, achieving in average the 83.75% of the process with a deviation of 22.65%. The last group, representing the 23.81% of the population, used version 39.0, getting an average of 88.72 with a deviation of 6.52.

### 5.5. Threats to validity

As omitting threats can lead to wrong conclusions, we present below some issues regarding the validity of our experiment design and the obtained results. First, we mention some factors that could represent a difference in the presented experimental instance:

- User experience. While none of the users knew about or ever used MoWA authoring tool, we have conducted the experiment with an heterogeneous group of participants with different capabilities concerning the usage of mobile devices, mobile browsers, browser extensions. We classify them by asking them some questions about their technological know-how. We collected results from 28.57% novice users, 28.57% regulars and 42.86% expert users. But even knowing their capabilities, we cannot predict a level of successfulness in the use of our tool.
- Learnability. Some users tend to learn the required tasks faster than others, and some of them require more practice or repetitions to achieve the same results. We had some repetitive tasks, like defining a marker for the PoIs, filling the required data and extracting data from external sites, placing and configuring augmenters.
- Interactions and understanding. Participants were allowed to interact during the experiment with others by face-to-face speaking (not performing the tasks instead), and we observed some of them did it at the beginning, leaving aside the instructions and asking to another participant. We cannot predict what would be happened with these few cases if the evaluation had been conducted individually.
- Maturation. As the users had to create 12 PoIs and the creation process took an average time of 1:07h, psychological changes can be experienced, as being fatigued, causing an impact in the successful tasks variable. They can also feel frustrated and quit. In fact, we had a case of a participant not ending the process.

As our experiment was a) controlled in a simulated environment where some conditions may differ in the real world, b) conducted against just one sample of participants and c) designed for solving a problem under just one application domain, we are aware that our findings can be altered when applied with different settings. Therefore, some setting can alter the representativeness of results, making our conclusions generalizable just under certain conditions. Below, we present some required points for experiment replications:

- Participant selection. We wanted a representative sample of end users to participate in our experiment, so a number of people was invited to participate in the experiment and they were the ones who decided volunteering in the experiment or not to do it. We did not discard any participant for any criteria, even if they were not used to mobile devices. In that sense, we ensured to have a heterogeneous group, however there were factors with prevailing values of people who chose to participate. The sample had: 71.43% of male members, 66,67% of Argentinean citizens, 90.48% of people with at least a degree career in course, 52.38% of coming from hard sciences field, 42.86% expert users, and aged not fall outside the range of 22 to 39 years.
- Context selection. We conducted the experiment in the facilities of our laboratory, covering four independent spaces, to ensure that the user has to move and look for the different points of interest.

However, the dimensions of the museum, the disposition and representation of the pieces of the collection were quite different. In addition, our facilities provided a stable Internet connection through WiFi.
- Domain selection and tasks complexity. As we wanted to compare our results against our previous results in [3], we decided to conduct the experiment under the unique, same domain problem, considering an only level of complexity.

## 5.6. Discussion

Our main purpose, the assessment of our research question, was successfully achieved favouring our approach; we demonstrated a high probability for an end user, without programming expertise, to be capable of creating his own mobile Web experience by reusing and enhancing an existing Web application with Context Aware features. Nevertheless, there are still many questions to answer, some of which derive from the following facts.

There was an unexpected behaviour during the experiment, concerning the single participant that quitted the experiment, completing a really low percentage of tasks; he achieved just a 26% of the process. He was in the right direction, but he told us he did not understand how to do it, and we could observe that unlike other participants, he did not attempt to solve the problem by watching how others did it, or interacting with them. He was part of the novice users' category, and yet, the results of this category of participants were promising.

There were also unexpected behaviours on the other side of the ledger. Participants with a low level of expertise were more successful in terms of percentage of completed tasks than regular users, and the difference against expert users was just of 7.34%. Among novices, the 83.3% never used Firefox for Android before, and none of them was used to use extensions for this browser. A 66.6% told us that he has never used any kind of browser extensions, and in a same percentage, they never used Web Augmentation applications or QR codes scanners. The 83.3% of them never used an authoring tool, and they never managed markers in a map. Many of these were challenges that the user had to face for the very first time to carry out the experiment, and anyway they obtained promising results in relation to experienced users.

Concerning the demographic and device data, and their relation with the level of completed tasks, we could observe that the higher values of standard deviation among the averages representing each category are related to the device platform (13.41%), the participant's level of studies (37.42%) and their area of interest (27.3%). This means that there was a greater degree of dispersion from the average value of the category. For example, while degree students (even the ones that abandoned the career), and degree and post-degree graduates achieved a level greater than 80%, the high school graduates' category has a 26.11% and the one for PhD students a 61.11%. However, drawing conclusions from these numbers might not be appropriate since four categories were represented by a single participant each (including the two of lowest results). In the case of the participant's area of interest, we can also observe that the ones belonging to Hard Sciences, Social Sciences, Naturals Sciences and Arts completed more than the 80% of the process, while the one from Economics did the 50% and the one with no interests achieved the 26.11%. Here, just one single participant represented these last two categories. Concerning the used device, lower values are related to the Pro Light (63.06) and Moto E (73.56%) devices. In the first case, we could not observe some striking difference against other platforms. Nevertheless, in the second case we observed that participants had a problem when sensing the QR codes, and the reason was that such phone model had a camera with fixed focus. This could influence both, the amount of work done and the participants' motivation. The lack of ability to focus made it hard the task of scanning QR codes, because at a close distance to it the captures were blurry and could not be properly interpreted by the decoding library.

Finally, we could also observe some complications related to the use of some UI elements. First, we used the XForms repeat object to create and list the properties of a Point of Interest, and many participants had some trouble interpreting this distribution in a small screen. A second problem we observed was related to an expected interaction by the users against the possible one. In concrete, we identified a problem in the interaction for inserting a new augmenter in the target Web page. As the contexts of the bar with the list of augmenters and the Web page content were different, and also because of the processing limitations of a mobile device, it was hard to simulate a drag and drop among both contexts. Instead, we implemented the insertion with a tap in the list of augmenters, and then, when the augmenter is inserted in the Web page context, the user was able to drag and drop the thumbnail.

## 6. Related Work

Having in mind that our focus is on End-User Development of Mobile Web Applications, we present below a series of related works facing EUD in the Mobile and Mobile Web fields. Along this section, we analyse each related work considering: the platform for which it was developed and the architecture of the tool; the resulting application kind;

the adopted end-user programming technique; the capability of the tool to solve different domain problems and to specify control flows, and the contemplation of crowdsourcing principles. Since the approaches are aimed at the end user, we also consider his expected expertise and the evaluation modality.

When contextual information is used to support the software development process, end users can build their solutions in the same place where the needs and problems are presented. This increases the boundaries of opportunistically development –a.k.a. situational applications [11]–. In this setting, [33] presents CASTOR, an authoring platform with a storytelling purpose. They implemented three supporting tools: a hybrid mobile application for creating stories in situ; a web-based client for managing the collected data, and a hybrid mobile application for consuming the narrations. The Phonegap framework supports the resultant applications, and users can share them through a repository. The authoring process is assisted through forms that allow users to select the structure of the story, defining stages and context values, writing the plot of the story. Stories can cover different domains –e.g. history, literature–, but they are always restricted to a storytelling purpose and using a limited number of story structures (simple, sequential or crossroad). The tool was evaluated with 19 primary school students with no programming skills, who were introduced with an initial briefing before usage.

Sometimes, users have needs that can be solved by combining existing Web content and services retrieved from different sources, and Mashups [16] may represent a possible solution. [13] present CAMUS, a framework for designing mobile applications through visual composition and high-level visual abstractions. It integrates and provides resources according to different contextual situations. It involves different user roles: an administrator with technical knowledge for registering resources in the platform and mapping them with context elements; a designer who defines how to mashup the services and to visualize the information; and a final recipient of the authored application. There is no mention of the adopted programming technique, but the provided screenshot suggests a WYSIWYG approach. CAMUS is aimed for creating context-aware mobile or Web-based applications. Nevertheless, the authors mention that execution engines are created as native applications for different mobile devices.

In the Mobile field, [38] presents MobiDev, an Android based development platform for mobile application creation from mobile devices. It allows the user to create the graphical interface by writing source code, designing mock-ups with a visual editor, or drawing a sketch on paper and taking it a picture, so the system will analyze and interpret it to generate a visual design. The approach contemplates end users with no programming skills developing applications with a basic control flow, but also developers defining a more specific behaviour through JavaScript code. The approach was evaluated with 16 students belonging to Computer Science department, who were previously trained. The experiment was successful, but the requirements did not contemplate the use of mobile features.

[21] presents a native platform for enabling end users to compose native mobile applications from their own mobile devices, by integrating the mobile features provided by the same device but also from Web services. Users specify activities through visualization components, which are executed by an interpreter that automatically creates the user interface. The approach comprises a repository for enabling end users to share their productions and re-configure them. The research team evaluated their approach with 40 students of the first year of the Computer Science career, and it was a requirement that they have no programming skills. Concerning the prior training, they give participants a lesson of 20 minutes about their tool, and another extra 20 about another similar tool against which they compared results.

Puzzle [15] is an EUD framework for producing native but Web-based applications, targeting touch-based platforms. Users combine building blocks through a puzzle-based metaphor with colour-equipped corners for pinpointing its combinability, which makes it suitable for end users with no programming skills. The authors mention that there is no need of plugins, but they end up presenting an implementation in the form of a native Android application. Diverse multi-purpose combinations can be created, changing the application's logic. A repository of created artefacts is available, but there is no way for users to request the construction of a concrete application to the crowd. Puzzle was evaluated with 13 participants with no IT-related jobs, and they were not exposed to previous training.

The authors in [6] propose an approach towards EUD for multi-device mashups creation with composite resources. They implemented a framework and a UI centric tool using the WYSIWYG technique. Users should select among the existing data components and UI templates in the repositories, and then perform an association between data items and visual elements. Finally, a platform-independent schema is generated and saved in the platform repository, so the user can download it in from the supported platforms and execute it through a native engine. Diverse multi-purpose combinations can be created and the design environment is a Web application, but they recommend not to execute it in mobile devices but in larger-screen ones. The experiment was conducted with a 10-minute demonstration and 36 participants, 17 of them with programming skills.

Other approaches empower the end users with the capability of creating Mobile Web applications from desktop environments. For instance, [30] allows the user to create widgets –in the meaning of simple applications– that represent a specific Web interaction. These artefacts, called Tasklets, are created using Programming By Example (PBE). Users need to install a plug-in in their desktop Web browser and use it to record the sequence of steps required to perform the task. This tool saves the need for representing these steps, builds a Tasklet template, detects and defines potential parameters, and finally makes the script accessible for multiple platforms through their repository. A wide spectrum of Tasklets could be created and shared for both, personal and public consumption.

Another related approach for EUD from desktop environments is presented in [31], where the authors present a cloud-based development platform of context-aware mobile services to be consumed as native applications. The platform is accessible through a web-based application, where the producer can associate a set of context values (specific locations, areas, times, dates, etc.) with concrete information to be delivered to the clients meeting such conditions. The authors conducted an experiment with 10 tourism domain-experts with no technical skills and no prior training, and they did it from a pre-installed native application. The resultant applications are bounded to the information delivery purpose.

Below, we present in Table 1 a summary of the presented approaches through 14 outstanding features. Analysed works are disposed as single rows, and features are arranged as columns. Applicant cases are ticked fields and unchecked values means a not supported or not mentioned feature in the corresponding writing.

|  | MOWA | [6] | [13] | [15] | [21] | [30] | [31] | [33] | [38] |
|---|---|---|---|---|---|---|---|---|---|
| Pure mobile development environment | ✗ | ✗ | ✗ | ✓ | ✓ | ✗ | ✗ | ✓ | ✓ |
| Pure mobile web development environment | ✓ | ✓ | ✗ | ✗ | ✗ | ✓ | ✗ | ✗ | ✗ |
| Produces pure mobile apps | ✗ | ✓ | ✓ | ✓ | ✓ | ✓ | ✓ | ✓ | ✓ |
| Produces pure mobile web apps | ✓ | ✗ | ✗ | ✗ | ✗ | ✗ | ✗ | ✗ | ✗ |
| End-user programming technique | Form-based WYSIWYG | WYSIWYG | WYSIWYG | Puzzle met. | Puzzle met. | PBE | Form-based | Form-based | Mockups |
| For multiple domains | ✓ | ✓ | ✓ | ✓ | ✓ | ✓ | ✗ | ✗ | ✓ |
| For non-technical users | ✓ | ✓ | ✓ | ✓ | ✓ | ✓ | ✓ | ✓ | ✓ |
| Users can request new requirements to the crowd | ✓ | ✗ | ✓ | ✗ | ✗ | ✗ | ✗ | ✗ | ✗ |
| Users can massively share their productions | ✓ | ✓ | ✓ | ✓ | ✓ | ✓ | ✓ | ✓ | ✗ |
| Users can specify the control flow | ✓ | ✓ | ✓ | ✓ | ✓ | ✓ | ✗ | ✗ | ✓ |
| Additional assistance mechanisms | ✓ | ✗ | ✗ | ✓ | ✓ | ✗ | ✗ | ✓ | ✓ |
| Authoring process at client-side | ✓ | ✓ | ✗ | ✗ | ✓ | ✗ | ✗ | ✗ | ✓ |
| Evaluation with non-technical users | ✓ | ✓ | ✗ | ✓ | ✓ | ✗ | ✓ | ✓ | ✗ |
| Evaluation without prior training | ✓ | ✓ | ✗ | ✓ | ✗ | ✗ | ✓ | ✓ | ✓ |

**Table 4:** Features covered by approach

As we can see, our approach faces the less common characteristics covered by existing works. The following ones are present in less than 50% of the works. A 37.5% of the solutions perform the authoring process at client-side; a 25% provide a pure mobile Web development environment; 12.5% offer end users the possibility to request new functionality to the crowd and there is no one allowing the creation of Mobile Web applications that do not rely on native components.

We also found interesting that the control flow definition feature is not contemplated by other approaches implemented through the same end-user programming technique than ours: the form-based one. In regard of such technique, we have overcomed multi-domain problems that the other form-based approaches do not. This is due to the domain-specific authoring tools inclusion based in building blocks combinations, so the difference is not achieved by the technique itself but through a combination of both in the whole development process.

**7. Concluding Remarks and Further work**

In this paper, we proposed an approach for empowering end users with the capability of creating Mobile Web applications from their mobile devices. Our solution consisted in a tool oriented to satisfy the end users' needs in multiple ways. In previous works, MoWA just contemplated the idea of a developer extending the framework and sharing applications with end users through a crowdsourcing platform. Therefore, end users were limited to ask for new implementations for a concrete scenario, and to download and install existing applications. Now, they can use our authoring tool for creating their own solutions for concrete scenarios.

Reusing artefacts of our previous framework also let us take advantage on the possibility of supporting an in-situ authoring process, from mobile devices themselves. The benefits of such feature, also in conjunction with live programming capabilities [32], are that the user is building his solution in the application's target execution context, whether virtual (a Web page) or real (a geographical position). He is dealing with the dimensions of the device, the modes of interaction, the layout of the elements in the DOM, the Internet connectivity, the physical presence of physical objects that could interfere when associating a geographic position with a digital resource, etc.

MoWA builders are created by developers who took technical decisions about the best way of solving domain-specific problems. Such tools supported the authoring process through a form-based implementation, assisting the user through a series of steps that allow them choose the best configuration for their scenario. It also made it possible to omit one step in the software development process where communication between people usually brings some understanding problems; requirements gathering. This step is now performed by the same person that will be on charge of the application creation. We are also enabling opportunistic development *in-situ*, as solutions are implemented for mobile platforms, being capable of accessing and using the mobile characteristics during the process.

By using our authoring tool, the end user's needs could be satisfied by themselves, and now they can play two roles in the Crowdsourcing platform: by demanding for concrete solutions –authoring tools/applications– and by solving other less-experienced user's needs. The worst possible scenario is an end user not finding a concrete sensor or builder, or not understanding how to combine components for a specific scenario problem. In this case, we are evaluating the benefits of enabling developers to create domain-specific authoring tools, so producers can use them for diverse but specialized concrete scenarios with a lower level of technical knowledge, because this way they do not need to take decisions about the proper sensor or kind of component to use in their application.

We demonstrated the feasibility of our approach by creating concrete builders for our authoring tool and using them for conducting an experiment with 21 participants, with a varied spectrum of demographic characteristics. Results proved that end users can deal with an *in-situ* creation process of Mobile Web Augmentation applications with a high level of confidence and 4-times faster than developers through scripting development. It also showed the feasibility of a tool overcoming client-side scripting from browsers extensions, with no native dependencies, and the devices' processing limitations.

Evaluation also allowed us to collect observations, ratings and opinions from the end users, which gave us the chance to improve the usability of the tools and the authoring process in general. From the technical point of view, we are also working to achieve resources optimization and implementing new ways of UI interactions, like drag-and-drop for the augmenter configuration thumbnails. New builders are being composed and we are also improving the augmenters' related ones for enabling the end users to specify new kind of interactions and execution order, so we can achieve a detailed application flow control.

We are also designing a DSL for applications specification, targeting different levels of knowledge and properly abstracting the required features. Our aim is to reach some common point between developers and end users, trying to generate new ways of collaborative development. In addition, we are also generating new mechanisms for supporting the development process in a collaborative environment and delegating some tasks to the crowd, with different levels of expertise. For example, an advanced end user can start a project in the community, by defining the basic structure of the application. Then, developers can contribute by enhancing the code with their low level technical knowledge, and some users can collect information about the real world, like pictures, locations, describing places in situ. Another participant can contribute in the project by collecting data through the Web for augmentation purposes. Users should be able to suggest changes and also to fork the whole project, if the visibility of the project is public, or private but shared with them. This will also lead us to analyse the evolution of projects and particularly the user's needs, behaviour and learning evolution.

**Disclosure Policy**

The authors declare that there is no conflict of interest regarding the publication of this paper.

**References**

1. Asakawa C. & Takagi H. Transcoding. In Web Accessibility (Springer London, 2008), pp. 231-260.
2. Baldauf, M., Dustdar, S., & Rosenberg, F. (2007). A survey on context-aware systems. International Journal of Ad Hoc and Ubiquitous Computing, 2(4), 263-277.
3. Bosetti, G.A., Firmenich, S., Gordillo, S.E. & Rossi, G. (2016). An approach for building Mobile Web applications through Web Augmentation. Journal on Web Engineering, In press.
4. Bouvin, N. O. (1999, February). Unifying strategies for Web augmentation. In Proceedings of the tenth ACM Conference on Hypertext and hypermedia: returning to our diverse roots: returning to our diverse roots (pp. 91-100). ACM.
5. Bouvin, N. O., Christensen, B. G., Grønbæk, K., & Hansen, F. A. (2003). HyCon: A framework for context-aware mobile hypermedia. New review of hypermedia and multimedia, 9(1), 59-88.
6. Cappiello, C., Matera, M., & Picozzi, M. (2015). A UI-Centric Approach for the End-User Development. ACM Transactions on the Web (TWEB), 9(3), 11.
7. Carlson, D., & Ruge, L. (2014, April). Ambient Amp: An open framework for dynamically augmenting legacy Websites with context-awareness. In Intelligent Sensors, Sensor Networks and Information Processing (ISSNIP), 2014 IEEE Ninth International Conference on (pp. 1-6). IEEE.
8. Challiol, C., Firmenich, S., Bosetti, G. A., Gordillo, S. E., & Rossi, G. (2013). Crowdsourcing Mobile Web Applications. In Current Trends in Web Engineering (pp. 223-237). Springer International Publishing.
9. Chang, S. K. (Ed.). (2012). Visual languages. Springer Science & Business Media.
10. Charland, A., & Leroux, B. (2011). Mobile application development: web vs. native. Communications of the ACM, 54(5), 49-53.
11. Cherbakov, L., Bravery, A., Goodman, B. D., Pandya, A., & Baggett, J. (2007). Changing the corporate IT development model: Tapping the power of grassroots computing. IBM Systems Journal, 46(4), 1-20.
12. Cheverst, K., Turner, H., Do, T., & Fitton, D. (2016). Supporting the consumption and co-authoring of locative media experiences for a rural village community: design and field trial evaluation of the SHARC2.0 framework. Multimedia Tools and Applications, 1-32.
13. Corvetta, F., Matera, M., Medana, R., Quintarelli, E., Rizzo, V., & Tanca, L. (2015). Designing and Developing Context-Aware Mobile Mashups: The CAMUS Approach. In Engineering the Web in the Big Data Era (pp. 651-654). Springer International Publishing.
14. Couthures, A. (2011). JSON for XForms. XML Prague 2011, 13.
15. Danado, J., & Paternò, F. (2012). Puzzle: a visual-based environment for end user development in touch-based mobile phones. In Human-centered software engineering (pp. 199-216). Springer Berlin Heidelberg.
16. Daniel, F., & Matera, M. (2014). Mashups: Concepts, models and architectures (1st ed.). Springer-Verlag Berlin Heidelberg.
17. Díaz, O., & Arellano, C. (2015). The Augmented Web: Rationales, Opportunities, and Challenges on Browser-Side Transcoding. ACM Transactions on the Web (TWEB), 9(2), 8.
18. Espada, J. P., Crespo, R. G., Martínez, O. S., G-Bustelo, B. C. P., & Lovelle, J. M. C. (2012). Extensible architecture for context-aware mobile web applications. Expert Systems with Applications, 39(10), 9686-9694.
19. Firmenich, S., Bosetti, G., Rossi, G., Winckler, M., & Barbieri, T. (2016, June). Abstracting and Structuring Web Contents for Supporting Personal Web Experiences. In International Conference on Web Engineering (pp. 77-95). Springer International Publishing.
20. Fling, B. (2009). Mobile design and development: Practical concepts and techniques for creating mobile sites and Web apps. O'Reilly Media, Inc.
21. Francese, R., Risi, M., Tortora, G., & Tucci, M. (2016). Visual Mobile Computing for Mobile End-Users. IEEE Transactions on Mobile Computing, 15(4), 1033-1046.
22. Gamma, E., Helm, R., Johnson, R., & Vlissides, J. (1994). Design patterns: elements of reusable object-oriented software. Pearson Education India.

23. Ghiani, G., Manca, M., Paternò, F., & Porta, C. (2014). Beyond responsive design: context-dependent multimodal augmentation of web applications. In Mobile Web Information Systems (pp. 71-85). Springer International Publishing.
24. Halbert, D. C. (1984). Programming by example (Doctoral dissertation, University of California, Berkeley).
25. Harper, S., Goble, C., & Pettitt, S. (2006). proximity: Walking the Link. Journal of Digital Information, 5(1).
26. Höllerer, T., & Feiner, S. (2004). Mobile augmented reality. Telegeoinformatics: Location-Based Computing and Services. Taylor and Francis Books Ltd., London, UK, 21.
27. Ko, A. J., Abraham, R., Beckwith, L., Blackwell, A., Burnett, M., Erwig, M. & Wiedenbeck, S. (2011). The state of the art in end-user software engineering. ACM Computing Surveys (CSUR), 43(3), 21.
28. Ko, A. J., Myers, B. A., & Aung, H. H. (2004, September). Six learning barriers in end-user programming systems. In Visual Languages and Human Centric Computing, 2004 IEEE Symposium on (pp. 199-206). IEEE.
29. Lieberman, H., Paternò, F., Klann, M., & Wulf, V. (2006). End-user development: An emerging paradigm (pp. 1-8). Springer Netherlands.
30. Manjunath, G., Murty, M. N., & Sitaram, D. (2013). Tasklets: enabling end user programming of web widgets. International Journal of Web Engineering and Technology, 8(3), 264-290.
31. Martín, D., Lamsfus, C., & Alzua-Sorzabal, A. (2016). A cloud-based platform to develop context-aware mobile applications by domain experts. Computer Standards & Interfaces, 44, 177-184.
32. McDirmid, S. (2007, October). Living it up with a live programming language. In ACM SIGPLAN Notices (Vol. 42, No. 10, pp. 623-638). ACM.
33. Pittarello, F., & Bertani, L. (2012). CASTOR: learning to create context-sensitive and emotionally engaging narrations in-situ. In Proceedings of the 11th International Conference on Interaction Design and Children (pp. 1-10). ACM.
34. Repenning, A., & Ioannidou, A. (2006). What makes end-user development tick? 13 design guidelines. In End User Development (pp. 51-85). Springer Netherlands.
35. Scaffidi, C., Shaw, M., & Myers, B. (2005, September). Estimating the numbers of end users and end user programmers. In Visual Languages and Human-Centric Computing, 2005 IEEE Symposium on (pp. 207-214). IEEE.
36. Schilit, W. N. (1995). A system architecture for context-aware mobile computing (Doctoral dissertation, Columbia University).
37. Schwinger, W., Grün, C., Pröll, B., Retschitzegger, W., & Schauerhuber, A. (2005). Context-awareness in mobile tourism guides–A comprehensive survey. Rapport Technique. Johannes Kepler University Linz.
38. Seifert, J., Pfleging, B., del Carmen Valderrama Bahamóndez, E., Hermes, M., Rukzio, E., & Schmidt, A. (2011, August). Mobidev: a tool for creating apps on mobile phones. In Proceedings of the 13th International Conference on Human Computer Interaction with Mobile Devices and Services (pp. 109-112). ACM.
39. Stolee, K. T., Elbaum, S., & Sarma, A. (2011, September). End-user programmers and their communities: An artefact-based analysis. In Empirical Software Engineering and Measurement (ESEM), 2011 International Symposium on (pp. 147-156). IEEE.
40. Van Woensel, W., Casteleyn, S., & De Troyer, O. (2011, June). A generic approach for on-the-fly adding of context-aware features to existing websites. In Proceedings of the 22nd ACM conference on Hypertext and hypermedia (pp. 143-152). ACM.